\documentstyle [12pt,epsfig]{article} 
\makeatletter

\@addtoreset{equation}{section}
\makeatother
\renewcommand{\theequation}{\thesection.\arabic{equation}}
\newcommand{\ba}{\begin{eqnarray}}
\newcommand{\ea}{\end{eqnarray}}

\newcommand{\TU}{T$_{\rm U}$ }
\newcommand{\TL}{T$_{\rm L}$ }
\newcommand{\GU}{G$_{\rm U}$ }
\newcommand{\GL}{G$_{\rm L}$ }
\newcommand{\RU}{R$_{\rm U}$ }
\newcommand{\RL}{R$_{\rm L}$ }
\newcommand{\YU}{Y$_{\rm U}$ }
\newcommand{\YL}{Y$_{\rm L}$ }
\newcommand{\DU}{D$_{\rm U}$ }
\newcommand{\DL}{D$_{\rm L}$ }
\newcommand{\DR}{D$_{\rm R}$ }
\newcommand{\DPR}{D'$_{\rm R}$ }
\newcommand{\DG}{D$_{\rm G}$ }
\newcommand{\DPG}{D'$_{\rm G}$ }

\begin{document}
\newcommand{\BS}{\bigskip}
\newcommand{\SECTION}[1]{\BS{\large\section{\bf #1}}}
\newcommand{\SUBSECTION}[1]{\BS{\large\subsection{\bf #1}}}
\newcommand{\SUBSUBSECTION}[1]{\BS{\large\subsubsection{\bf #1}}}

\begin{titlepage}
%\hspace*{8cm} {UGVA-DPNC 1998/04-176 April 1998}
\hspace*{8cm} {arXiv:physics/0606135 v1, June 2006}
\begin{center}
\vspace*{2cm}
{\large \bf The train/embankment thought experiment, Einstein's second postulate
 of special relativity and relativity of simultaneity}  
\vspace*{1.5cm}
\end{center}
\begin{center}
{\bf J.H.Field }
\end{center}
\begin{center}
{ 
D\'{e}partement de Physique Nucl\'{e}aire et Corpusculaire
 Universit\'{e} de Gen\`{e}ve . 24, quai Ernest-Ansermet
 CH-1211 Gen\`{e}ve 4.
}
\newline
\newline
   E-mail: john.field@cern.ch
\end{center}
\vspace*{2cm}
\begin{abstract}
  The relativistic time dilatation effect and Einstein's second postulate of special
  relativity are used to analyse train/embankment thought expriments, both Einstein's
  original one, and an extension where observers on two trains moving at different speeds,
   as well as on the embankment, are considered. Whereas the relativistic analysis
   of Einstein's experiment shows, in contradiction to Einstein's interpretation,
   no `relativity of simultaneity' effect, the latter is apparent for certain events
   in the two-train experiment. The importance of relativistic kinematics ---embodied
   for photons in Einstein's second postulate--- for the correct interpretation of
   the experiments is pointed out and demonstrated by detailed calculation of
   a related example.
 \par \underline{PACS 03.30.+p}
\vspace*{1cm}
%\par Published in: {\it American Journal of Physics {\bf 68} (2000), 267-274.} 
\end{abstract}
%\vspace*{4cm}cp
%24 pages, 14 figures, 7 tables
\end{titlepage}
%Space Time Measurements...
%\par J.H.Field,D\'{e}partement de Physique Nucl\'{e}aire et Corpusculaire
% Universit\'{e} de Gen\`{e}ve~, 24, quai Ernest-Ansermet
% CH-1211 Gen\`{e}ve 4. e-mail: john.field@cern.ch. Tel: (41) 22 702 6274/6273.
% Fax: (41) 22 781 21 92   
 
\SECTION{\bf{Introduction}}
 Einstein's Train-Embankment Thought Experiment (TETE), introduced in his popular
 book `Relativity, the Special and General Theory'~\cite{EinSGR}, first published
 in 1920, has been used in many introductory textbooks\footnote{For example, Refs~\cite{WT,TL}.
  See also numerous citations in Refs~\cite{Nelson,GEB}.}
 and articles in the pedagogical
 literature~\cite{SSV}, to introduce the concept of `relativity of simultaneity'
 of Special Relativity (SR) before considering the space-time Lorentz Transformation (LT).
  \par Since the complete relativistic interpretation of the TETE depends both on
   direct physical consequences of the space-time LT ---the time dilatation effect--- and an
   understanding of the
   relativistic kinematics of photons as embodied in Einstein's second postulate of SR, the possibility
   of introducing  `relativity of simultaneity' in this simple way is illusory. This will
   become clear later in the present paper. However, as will be shown, a somewhat more
    sophisticated TETE involving observers on the embankment and in two trains, with suitably
    chosen speeds, does indeed demonstrate the existence of a genuine
    relativity of simultaneity effect for certain events. In contrast, the correct
     relativistic analysis of Einstein's original TETE shows that the `lightning strokes'
     will be judged to be simultaneous by both the train and the embankment observers,
     in contradiction to Einstein's conclusion.
     \par The fallacy in Einstein's reasoning is simple to understand. If the experiment
     were to be performed with non-relativistic massive particles of constant velocity replacing
     photons, an analysis of the various space-time events could be performed in either
    the embankment or the train frames and would produce identical results. In the case of
   photons or massive relativistic particles, as shown explicitly in the example discussed
   in the Appendix, this is no longer the case. Then a correct analysis of events in the
   train frame requires that such events be specifically considered. This was not done in Einstein's
   interpretation, where only events observed in the embankment frame were considered.
    \par The structure of the paper is as follows: In the following section, after a brief
    discussion of the simplest axioms which may serve as the basis for SR, two consequences
    of SR ---Invariance of Contiguity (IC) and the time dilatation (TD) effect--- are 
     derived. Application of IC, TD and Einstein's second postulate of SR are 
    sufficient for complete analyses of the TETEs discussed in later sections of the paper.
    In particlar, explicit use of the space-time LT is not required. Section 3 presents
    Einstein's original TETE and discusses it, and Einstein's interpretation of it, in 
    a critical manner. A similar experiment where photons are replaced by sound signals,
    either in the atmosphere or in the interior of the train, is also discussed.
    In Section 4 a more elaborate TETE with two trains and a well-defined procedure
    for synchronous production of light signals is described and analysed. It is shown
    that events corresponding to light signals, triggered by coincidence detection of
    two other light signals in different inertial frames, show a genuine
    relativity of simultaneity effect quite distinct from the one proposed by Einstein.
    Section 5 contains a discussion of the distinction between `relative velocity'
    and `speed' as applied in the TETE as well as the closely related operational meaning
    of Einstein's second postulate. The latter is also shown to be a direct consequence
    of the relativistic kinematics of massless particles~\cite{JHFLT}. Section 6 contains
    a detailed discussion of a recent paper by Nelson~\cite{Nelson} that gives a 
     re-interpretation of Einstein's TETE. Nelson finally makes the same fundamental
    mistake, mentioned above, of attempting an analysis of the problem purely in terms
     of embankment frame events, although earlier in the paper came close to the correct
    solution of the problem. This was rejected due to a misapplication of IC to
    different events (falsely assumed to be the same) in the two frames. A summary and the conclusions
   of the present paper are to be found in Section 7. An appendix contains an analysis of the TETE from
    a different point of view. Instead of analysing light signals produced by the lightning
    strokes, the latter are replaced by pulsed laser beams in the embankment frame
     and the question of the simultaneity or non-simultaneity of the events where the laser
    pulses strike either the embankment at points aligned with the ends of the moving train,
    or the train itself,
    is considered in both the embankment and train frames. The results obtained demonstrate
    immediately the fallacy of Einstein's embankment-frame-only interpretation of the TETE.
    \par Previous papers by the present author have pointed out the spurious nature
   of the correlated
     `Length Contraction' (LC) and `Relativity of Simultaneity' (RS) effects derived by
     misapplication of the space-time LT~\cite{JHF1,JHF2,JHF3,JHF4}. These effects were
    invoked in Nelson's final `relativistic' interpretation of the TETE. The reader is
    referred to these papers
    for a critique of the conventional interpretation of the space-time LT, and particularly
    to Ref~\cite{JHF3} in which the essential arguments are concisely presented. However these
    arguments are also given in the present paper in Section 4 (Eqn(4.2) and Section 6
     (Eqns(6.20)-(6.12)). Recognition of the spurious nature of these LT related RS and LC
    effects is essential to fully understand the relativistic analyses of TETEs presented
    in this paper.

\SECTION{\bf{Axiomatic bases of special relativity. Invariance of contiguity}}

%\begin{figure}[htbp]
%\begin{center}\hspace*{-0.5cm}\mbox{
%\epsfysize15.0cm\epsffile{stimef1c.eps}}
%\caption{ {\em  The caption.}}
%\label{fig-fig1}
%\end{center}
%\end{figure}

  Einstein's original formulation of special relativity (SR) was founded on two main postulates
 \footnote{Einstein stated explicitly only the two postulates E1 and E2 in
     Ref.~\cite{Ein1}, but at least three other postulates: linearity of the equations,
      spatial isotropy, and the reciprocity postulate (see below) were tacitly
     assumed.}
 \par (E1) {\tt The dynamical laws of nature are the same in any inertial frame.} 
 \par (E2) {\tt The speed of light is the same in any inertial frame, and is \newline independent 
     of whether the source is stationary or in motion.}
   \par One aim of the present paper is to discuss the precise operational meaning of the
   postulate E2; however, as was realised shortly after the advent of SR~\cite{Ignatowsky},
    the space-time LT, from which all predictions of SR may be obtained, can be
   derived without invoking E2. See~\cite{BG,JHFLT} for surveys of the related literature.
   For example, in Ref~\cite{JHFLT}, the LT is derived from the requirement that it is a
   single-valued function of the space-time coordinates and the following postulate,
   that is a weak operational statement of the relativity principle:
 \par (MRP) {\tt Reciprocal measurements of similar rulers and clocks at rest in \newline two different
     inertial frames, by observers at rest in these frames, yield \newline identical
     results.}
  \par  It will be instructive, in the present paper, in discussing events observed from
    different inertial frames,
    to check that this Measurement Reciprocity Postulate (MRP) is respected in each case.
   Unlike in E1, no dynamical laws are in invoked by the MRP. Another general property of
    SR that will be found of great importance when comparing different interpretations
    of the TETE is Invariance of Contiguity (IC) which may be derived directly from 
    the LT:
 \par (IC) {\tt Two events which are in space time contiguity in one inertial frame
            are so in all inertial frames.} 
      \par  Pairs of space-time contigouus events may be, in an obvious way, `$xt-$', `$xyt-$',
        or `$xyzt-$ contiguous'. In the following it will be sufficient to consider
    `$xt-$contiguous' events. The proof of IC is simple.
      The space-time LT relating the space-time coordinates of events as specified by
      synchronised clocks in two inertial frames S and S' is:
  \begin{eqnarray}
     x' & = & \gamma_v(x-vt) \\
   t' & = & \gamma_v(t-\beta_v x/c) \\
      y' & = & y  \\
      z' & = & z
  \end{eqnarray}
   where $\beta_v \equiv v/c$, and $\gamma_v \equiv 1/\sqrt{1-\beta_v^2}$. 
     In (2.1)-(2.4) the frame S' moves with speed $v$ along the common $x$, $x'$ axis of S and S'.
     This usual form of the LT, as derived by Einstein, is valid provided that the 
     space time origin in S' is coincident with the $x'$ coordinate of the transformed
    event. The appropriate LT when this condition is not satisifed is discussed in 
    Section 6 below. 
    Any two events in S: ($x_1$,$y_1$,$z_1$,$t_1$) and ($x_2$,$y_2$,$z_2$,$t_2$)
    and the corresponding events in S':($x'_1$,$y'_1$,$z'_1$,$t'_1$) and
    ($x'_2$,$y'_2$,$z'_2$,$t'_2$) that satisfy (2.1)-(2.4) also respect the 
     interval relations, that are independent of the choice of the origins
    of coordinate systems in S and S':
  \begin{eqnarray}
     x'_1-x'_2 & = & \gamma_v[x_1-x_2-v(t_1-t_2)] \\
   t'_1-t'_2 & = & \gamma_v[t_1-t_2-\beta_v(x_1-x_2)/c] \\
      y'_1 - y'_2 & = & y_1-y_2 \\
      z'_1-z'_2 & = & z_1-z_2
  \end{eqnarray}
   If events 1 and 2 are  `$xt-$contiguous' in S. i.e. if $x_1 = x_2$ and $t_1 = t_2$
   It follows from (2.5) and (2.6) that also  $x'_1 = x'_2$ and $t'_1 = t'_2$. Thus the
   events are contiguous in S' for any value of $v$, that is, in any inertial
   frame. An immediate and important corollary for the discussion of the TETE is that if 
    the $x$-$t$  world lines of three physical objects intersect at a common
    point in any inertial frame, i.e. if the three events at the intersection are $xt-$contiguous,
    they must intersect at a common point in any inertial frame.
    \par This demonstration is the only direct use of the space-time LT in
     following discussion of TETEs. It will be found possible to completely
     analyse the different thought experiments considered by noting that a light-transit
   time in a given inertial frame between objects at rest in that frame, constitutes
   a `photon clock'~\cite{RPFPC} that measures the proper time in that frame.
   As do all clocks when in uniform motion, such clocks display the time dilatation (TD)
   effect.
       The prediction of the TD effect relating a time interval, $\Delta t'$, recorded by a clock
   at rest in the frame S', but as observed from the frame S , 
   to corresponding time interval, $\Delta \tau$, as measured by a clock at rest in S,
   may be derived from the time-like invariant interval relation:
   \begin{equation}
     c^2(\Delta \tau)^2-(\Delta x)^2 =  c^2(\Delta t')^2-(\Delta x')^2
   \end{equation}
   A clock at a fixed position in S' has $\Delta x' =0$. Since also $\Delta x = v \Delta \tau$
   it follows from (2.9) that:
    \begin{equation}
     c^2(\Delta \tau)^2(1-\frac{v^2}{c^2})  = c^2 \frac{(\Delta \tau)^2}{\gamma_v^2} =  c^2(\Delta t')^2
   \end{equation}
    or
      \begin{equation}   
    \Delta \tau = \gamma_v \Delta t'
     \end{equation}
   Consideration of a clock at rest in S, but observed from S', leads to the similar formula:
      \begin{equation}   
    \Delta \tau' = \gamma_v \Delta t
     \end{equation}    
    where $\tau'$ is the time recorded by a clock at rest in S', and $\Delta t$ is the
     corresponding time recorded by a similar clock at rest in S but observed from  S'. 
    \par It is very important to note, in the analysis of space-time problems in special relativity,
     the existence of the four different time intervals  $\Delta \tau$,  $\Delta \tau'$,
     $\Delta t$ and $\Delta t'$. Where  $\tau$ and $\tau'$ are observed proper times and
      $t$ and $t'$ are the apparent times of moving clocks. The  intervals  $\Delta \tau$
     and  $\Delta t'$ relate to one possible experiment and $\Delta \tau'$ and  $\Delta t$ to
     a different, although reciprocal, in the sense of the MRP, one. Thus when the LT are
      used to analyse experiments it  always gives a relation between a proper time and
     an apparent time, never between two proper or two apparent times. The erroneous identification
     of $\tau$ with $t$ or of  $\tau'$ with $t'$ has resulted in many incorrect predictions
      of relativistic effects, particularly in classical electrodynamics~\cite{JHFFT}. 
      Another example of this error of interpretation of the LT is provided by Nelson's analysis of
      the TETE~\cite{Nelson}
     to be discussed below in Section 6. Indeed, the logical absurdity of setting
     $\Delta \tau = \Delta t$ and  $\Delta \tau' = \Delta t'$ thus mixing up the time intervals
      recorded in different experiments becomes evident on subsituting these relations
      into (2.11) and (2.12) to yield the equations:
     \begin{equation}   
    \Delta \tau = \gamma_v \Delta \tau'
     \end{equation}
    \begin{equation}   
    \Delta \tau' = \gamma_v \Delta \tau
     \end{equation} 
   Substituting (2.14) into (2.13) gives $\gamma_v^2 = 1$ or $v = 0$, whereas $\gamma_v$ in (2.11) and (2.12)
   can take any positive real value. On the other hand, the substitutions $\Delta t \leftrightarrow \Delta t'$
    and $\Delta \tau \leftrightarrow \Delta \tau'$ correspond to the valid operation
    of replacing a time dilatation
    experiment by a reciprocal one, which has the effect of exchanging Eqns(2.11) and (2.12).
   \par The convention of using the
   symbol $\tau$ to denote the observed time of a clock at rest and $t$ to denote the
    observed time of a clock in uniform motion is followed throughout the present paper.
   \par Eqns(2.11) and (2.12) may also be written in a more explicit notation as:
    \begin{equation}   
    \Delta \tau_{rest} = \gamma_v \Delta \tau'_{move}
     \end{equation}
    \begin{equation}   
    \Delta \tau'_{rest} = \gamma_v \Delta \tau_{move}
     \end{equation}
   where all intervals refer to proper times and the subscripts specify whether the corresponding
   clock is in motion or at rest. It is this form of the TD interval relation that is used to
   extract the proper lifetime of an unstable particle $\Delta \tau'_{move}$ from the measurement
    of the corresponding laboratory lifetime  $ \Delta \tau_{rest}$ knowing the value
  of the relativistic parameter $\gamma_v$ from kinematics or direct velocity
   measurement. 
\SECTION{\bf{Einstein's presentation of the TETE}}
\begin{figure}[htbp]
\begin{center}\hspace*{-0.5cm}\mbox{
\epsfysize15.0cm\epsffile{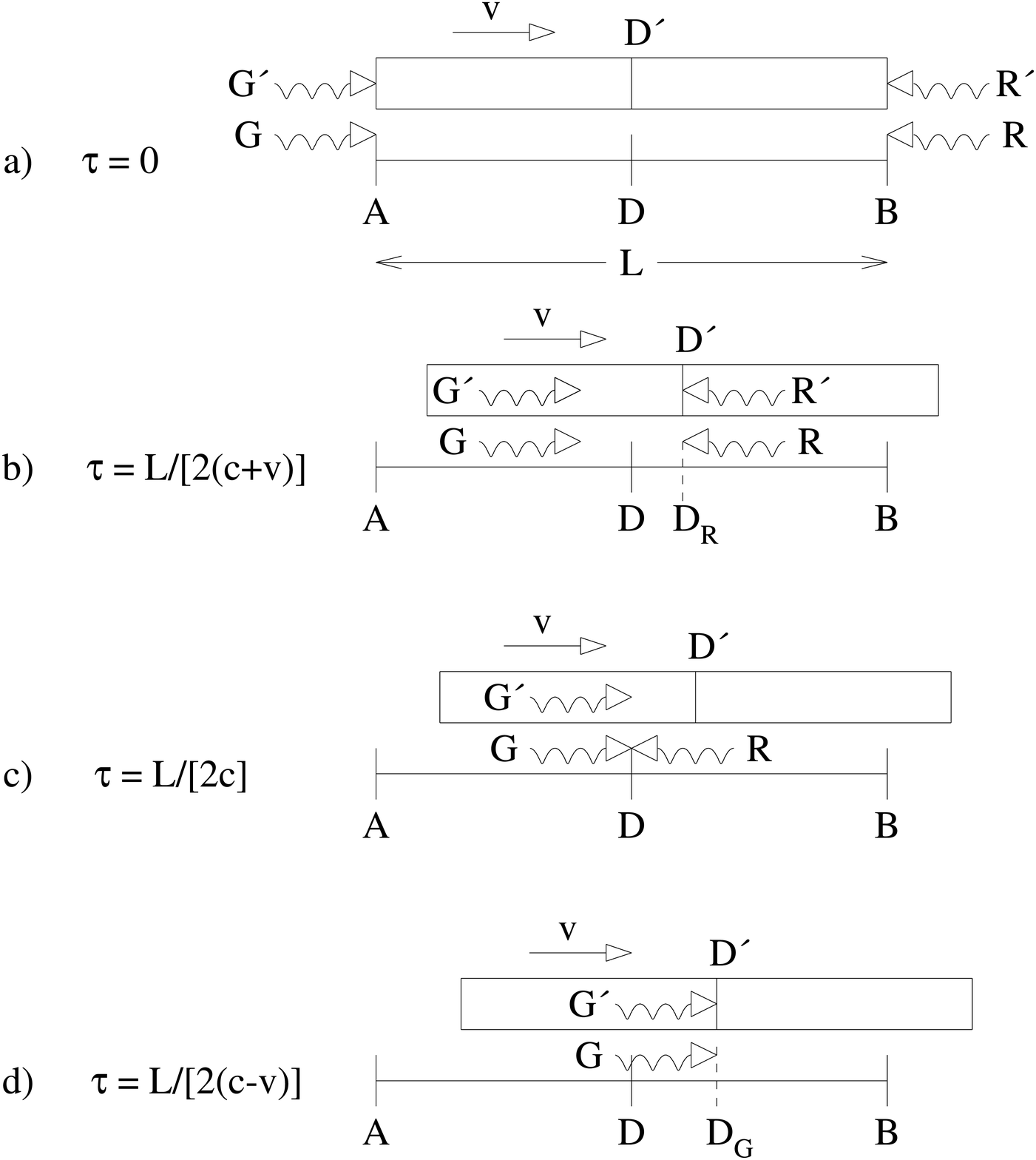}}
\caption{ {\em Einstein's interpretation of the TETE in the embankment
    frame. a) $\tau = 0$, photons G',G, R' and R are emitted from lightning stroke events at
     A and B. b) $\tau = L/[2(c+v)]$ , photons, R, R' are spatially contiguous with \DR and D'
    respectively. c) $\tau = L/[2c]$ photons G and R are detected in coincidence by D.
     d) $\tau = L/[2(c-v)]$,  photons G, G' are spatially contiguous with \DG and
     D' respectively. In this figure $v = c/4$.  The photons G' and R' that are detected by D' in
    the train are 
     assumed to be $xt$-contigous with the photons G and R, respectively, that are 
    detected by D on the embankment. Since the analysis is performed entirely in the embankment
    frame the events R'-D'; and G'-D'considered by Einstein are actually the embankment frame
    events R-\DR
    and G-\DG, not events in the train frame. Thus the essential problem is not addressed.
     In order to discuss the simultaneity (or lack of it) of events in the train frame, an 
   analysis of such events, in the train frame itself, must be performed. As discussed in the Appendix,
   if the photons are replaced by
    non-relativistic massive particles analyses in the train and embankment frames are equivalent,
    so that a correct space-time analysis may be performed in either frame. This is no 
    longer the case in special relativity, invalidating Einstein's interpretation.}}
\label{fig-fig1}
\end{center}
\end{figure}

\begin{figure}[htbp]
\begin{center}\hspace*{-0.5cm}\mbox{
\epsfysize15.0cm\epsffile{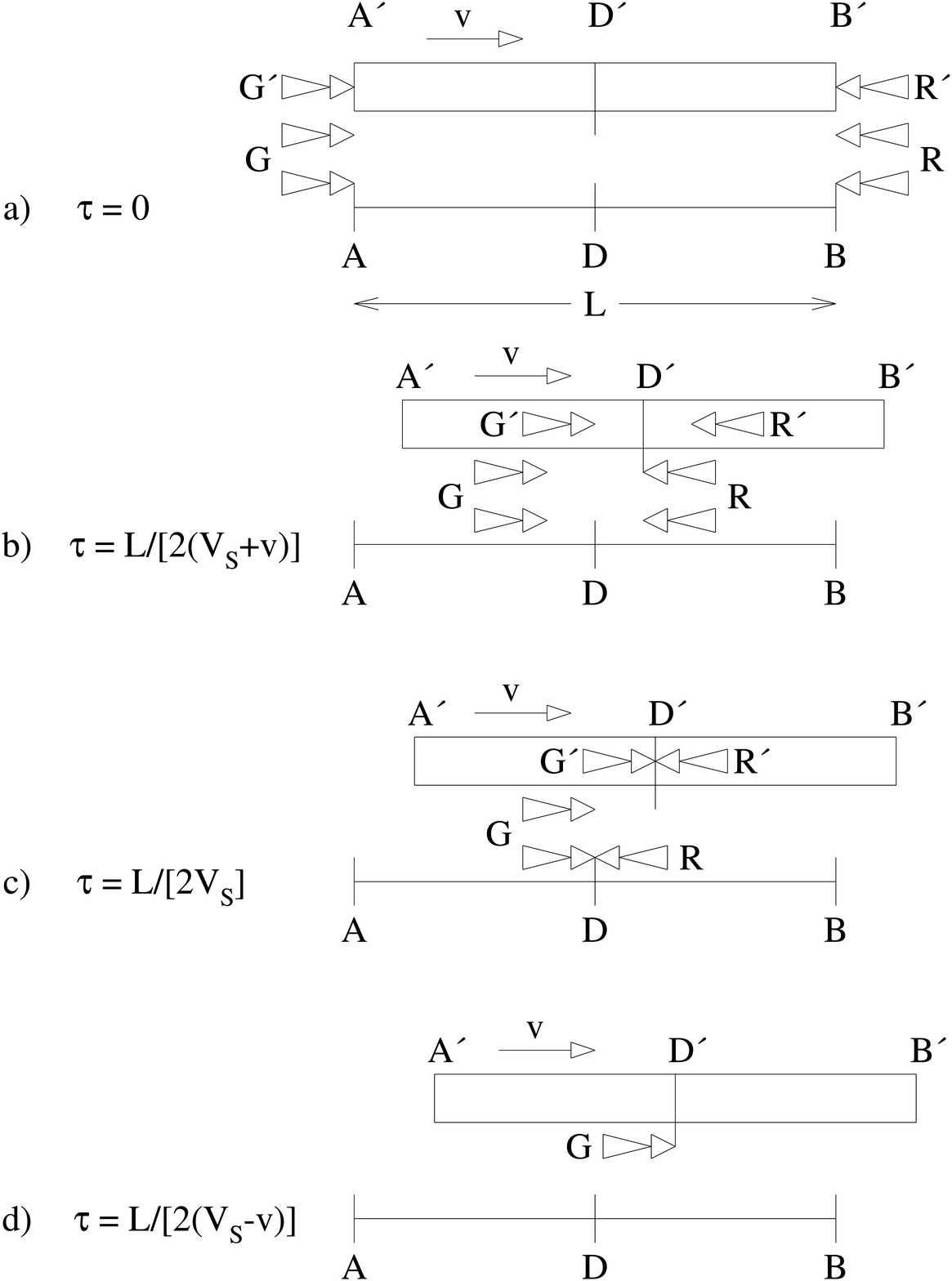}}
\caption{ {\em  The Einstein TETE with light signals replaced by sound signals. If uniform sound
   propagation, in the stationary atmosphere, unperturbed by the motion of the train, is considered
    (signals G, R) results strictly analgous to Einstein's for the case of light signals are 
    obtained with the replacement $c \rightarrow V_S$. If the sound signals in the train instead move 
   in air that is at rest relative to the
     train, (signals G',R'), the signals are observed simultaneously in the train [ c)].
    In this figure $v = V_S/4$. In view of
    Einstein's second postulate, replacing the speed $V_S$ of sound by the speed of light $c$, it may
    be expected that the light signals would also be observed simultaneously in the train, 
    in contradiction to Einstein's conclusion.}}
\label{fig-fig2}
\end{center}
\end{figure}

\begin{figure}[htbp]
\begin{center}\hspace*{-0.5cm}\mbox{
\epsfysize15.0cm\epsffile{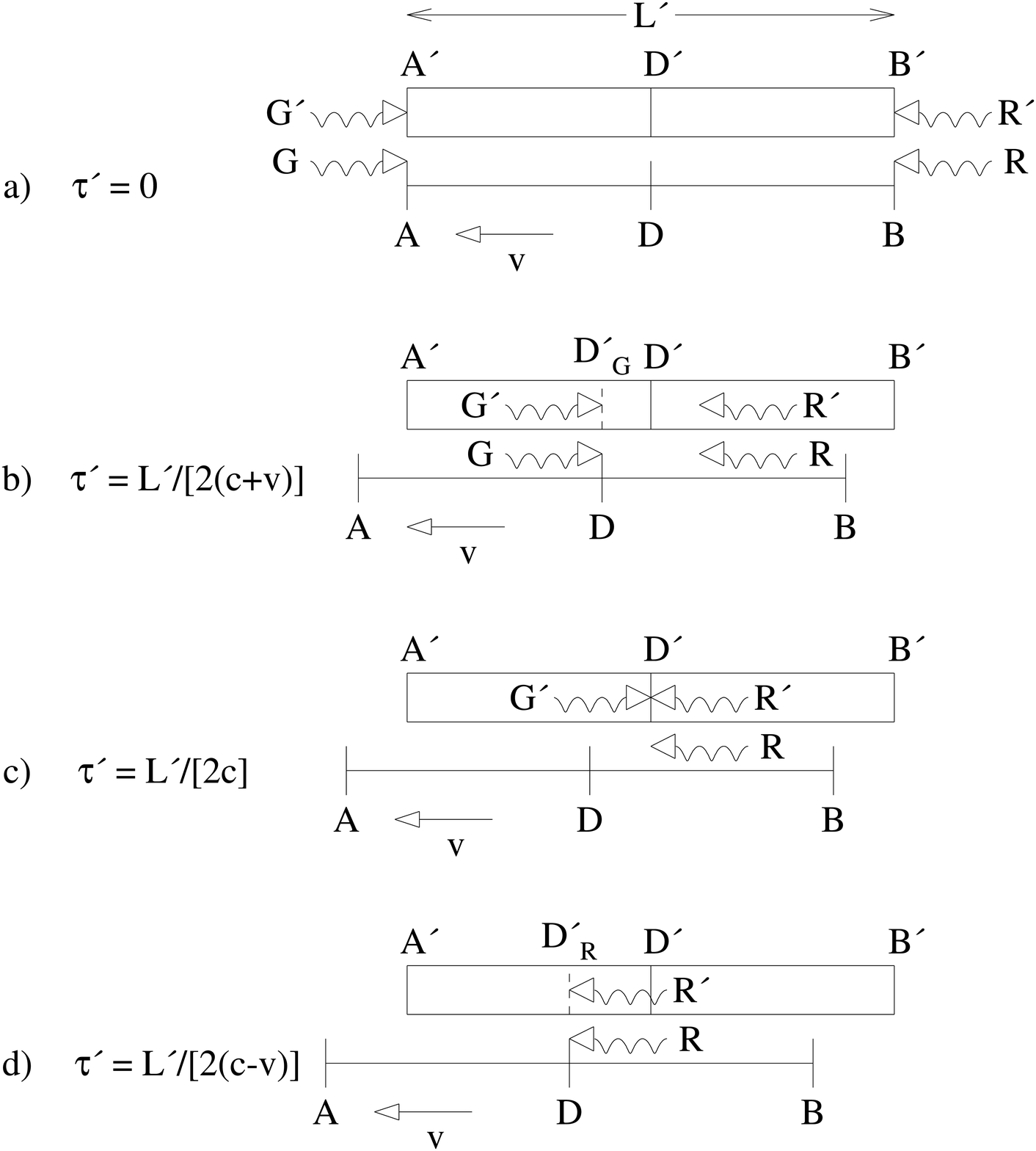}}
\caption{ {\em  Einstein's interpretation of the TETE in the train frame. As in Fig.1, $v = c/4$.
    It is assumed that
   the lightning stroke events are also simultaneous in this frame. As shown in the text, 
   this is the case in both Galilean and special relativity. The events \newline G'-\DPG, G'-R'-D'
    and R'-\mbox{\DPR} are specified in the train frame. The times of the events are obtained
    from those
   shown in Fig.1 by exchange of primed and unprimed quantities, so that the experiments
    shown in this figure and Fig.1 are `reciprocal' ones. In Einstein's interpretation
    the events G'-\DPG and R'-\mbox{\DPR} in the train frame are wrongly identified  with
   putative embankment frame events G-D and R-D respectively.}}
\label{fig-fig3}
\end{center}
\end{figure}

  In his original presentation of the TETE~\cite{EinSGR}, Einstein considered a train 
  moving with uniform velocity, $v$, along an embankment. In the following, the inertial frame
 of the embankment is denoted by S, that of a train  by S' or S''. The positive $x$-axis is
 parallel to the direction of motion of the train. Two lightning strokes hit points A and B of
 the embankment, simultaneously in S. The spatially contiguous points on the train, at the instant
 of the lightning strokes are denoted by A' and B'. It is asked whether the lightning strokes
 will be perceived by an observer on the train as simultaneous, or not simultaneous. To answer this
 question, observers are introduced at M and M' on the embankment and the train respectively. M is midway
 between A and B and M' is midway between A' and B'. The production process of the light signals 
 in the frames S and S' associated with the lightning strokes and observed at M and M' was not
  discussed by Einstein. As will be seen, this is not important for Einstein's interpretation
  since no observations performed in the frame S' are considered.
  \par Einstein's analysis of the TETE is presented in Fig.1. For definiteness, the `observers'
  at M and M' are replaced by photon detectors D and D'. Using his light signal synchronisation
  procedure, Einstein argues that the events at A and B (or at A' and B') will be simultaneous
  provided that the corresponding light signals, as observed in the embankment frame,
    are simultaneous at D (or at D'). From this, it is concluded that the photon (or light signal)
   initiating events are simultaneous for the observer on the embankment, but not for the
    observer on the train. Clearly (assuming all photons have the same speed in a given
   inertial frame) the events will be judged to be simultaneous by the observer on the
    embankment (see Fig.1c). Einstein gives the following argument concerning the observation
   of the light signals on the train~\cite{EinTETE}:
    \par {\tt Now in reality (considered with reference to the railway embankment)
      he} (i.e. the observer in the train) {\tt is hastening towards the beam of light coming
     from B, while he is riding on ahead of the
     beam of light coming from A. Hence the observer will see the beam of light emitted
      by B earlier than that emitted from A. Observers who take the railway train as
      the reference-body must \newline therefore come to the conclusion that the lightning flash at
      B took place \newline earlier than the lightning flash at A. We thus arrive at the important
      result:\newline Events which are simultaneous with reference to the embankment are not
    \newline simultaneous with respect to the train and vice versa (relativity of
     \newline simultaneity).}
      \par Actually, however, all the observations described in this passage are performed
        by observers in the embankment frame, never by an observer on the train.
        When the photon R' arrives at D' (Fig.1b) before the photon G'
        arrives at D' (Fig.1d) what are observed are events in S, not in S'.
        To specify precisely, in an operational manner, these events which are observed in S,
        it is convenient to intoduce the detectors \DR and \DG, at rest in S, that are spatially contiguous
        with D' at the apparent instants of photon coincidences with the latter detector
        (see Fig.1b and 1d).
          Since these clocks are at rest in S, a definite time (the proper time in S) can be assigned
       to the R-\DR and G-\DG coincidence events. This is not the case for the
       time in S' corresponding to apparent R'-D' and G'-D'
       coincidences. Since the detector D' is in motion, the time of the photon detection events in S',
        which is what is essential for problem under consideration, cannot be
       {\it a priori} identified with the S-frame times
     of the events  R-\DR and G-\DG, which is what is assumed in Einstein's interpretation. So, contrary to
            Einstein's statement, no conclusion can be drawn concerning the simultaneity,
        or non-simultaneity of the observation of the photons R' and G' in the frame S'.
        Indeed, the postulate E2, stating the equality of the speed of light in the frames
        S and S', is not invoked in Einstein's argument, as it must be (and is in the
       following section) in order to obtain the correct prediction of SR concerning the simultaneity
       (or lack of it) of the photon detection events observed in S or S'. For this it is
       necessary to consider events not only in the frame of the embankment, as Einstein 
       does, but also events in the frame of the train. 
       \par The argument given in Einstein's passage quoted above is very similar
         to that presented in the original paper on SR~\cite{Ein1} where `relativity
         of simultaneity' was first introduced. A `photon clock' in motion was compared
         with one at rest, but only events in a single inertial reference frame were
         discussed. Indeed, at this point in Ref.~\cite{Ein1}, no meaningful discussion
        of relativity of simultaneity was possible because the LT had yet to be derived.
        Similarly in Ref.~\cite{EinSGR}, the LT necessary for a correct analysis of
         relativity of simultaneity is introduced only in Chapter XI after previously
         introducing the TETE in Chapter IX and relativity of distance (also based on the TETE)
         in Chapter X.
        \par In fact, the sequence of events seen in a single inertial frame shown in Fig.1 would
         be identical in SR, where postulate E2 applies, or in the preferred frame (aether rest frame)
        of a theory where E2 does not apply, but instead light may have different speeds in
        different inertial frames, in the case that the embankment is associated with this
         preferred frame. Indeed, as shown in Fig.2, the pattern of events is the same
         if the photons in Fig.1 are replaced by sound signals in air, if the train does not
         drag the air in the regions where the sound signals are propagating. If however the sound
         signals were detected {\it inside} the train (assuming that the  centre of mass of the
         air in the train is rest in S') the production events would be judged simultaneous by
         both the train and embankment observers (Fig.2c). However, the sound signals 
         in S would not be seen by an observer in this frame to arrive simultaneously
         at the {\it position of} D' {\it in} S (Fig.2b and 2d). In the limit where the speed of sound
         in air, $V_S$, is much less than the speed of light, $c$, not only will the R and G (R' and G')
         signals be received simultanously by the detectors D (D') but also the G'-R'-D'
        and G-D-R coincidence events will be seen to be simultaneous by observers in both
         S and S' (Fig.2c). As will be seen in the following section, this is no longer the case 
        when corrections of O($(V_S/c)^2$) are included.
         \par Since the postulate E2 states that the speed of light signals is the same in
        S and S', regardless of the motion of the source, it would seem that the situation shown in
         Fig.2 for the motion of the sound signals in the train should be strictly analogous to that
         of the light signals in vacuum inside the train, replacing $V_S$ by $c$. The physics underlying
        the propagation of the signals is quite different in the two cases (constant speed of sound
        relative to air in one case, Einstein's postulate E2 in the other) but, in both cases, the
       signals would be expected, contrary to Einstein's conclusion, to be simultaneous, both in the
       train and on the embankment.
        \par Einstein chose to analyse the problem entirely in the reference frame of the embankment
         observer. What happens if it is instead analysed in the reference frame of the observer on
         the train?  This is shown in Fig.3, where it is assumed that the lightning stroke
        events are also simultaneous as seen by local observers\footnote{i.e., observers situated 
       at A' and B'.} in the train frame. The validity of this assumption
         is further discussed below, as well as in the Appendix.
         Performing exactly the
        same analysis as Einstein, but instead considering events only in the frame S' instead
       of only in the frame S, it would be concluded, from Einstein's argument, that the production
       events of the photons G' and R' are simultaneous (Fig.3c) whereas those of G and R  are not
      (Fig.3b and Fig.3d). This different behaviour is perfectly compatible with the relativity
       principle as embodied in the MRP. However, there is an apparent violation of
       invariance of contiguity if it is assumed that the G-R-D coincidence in Fig.1 and 
        the G-D and R-D coincidences in Fig.3 represent the same event.  
       The $x$-$t$  world lines of G, R and D intersect in a point in S (Fig 1c) but not in S'
       (Fig.3b and Fig.3d) Similarly the $x'$- $t'$ world lines of  G', R' and D' intersect in
       a point in S' (Fig.3c), but not in S (Fig.1b and Fig.1d). Thus it appears that Einstein's claimed
       `relativity of simultaneity' effect is incompatible with IC, that is, as shown in the previous
        section, a necessary consequence of the LT.  However, as discussed above, such
        an identification of all photon detection events as common events, viewed either in S or in S',
      is fallacious The non-simultaneous
     events in S are R-\DR and G-\DG not R'-D' and G'-D' respectively, while the observed events in
       S' are (see Fig.3) R'-\DPR and G'-\DPG not R-D and G-D in S. 
     Indeed, R-\DR, G-\DG, R-G-D,  R'-\DPR, G'-\DPG
           and R'-G'-D' are all physically distinct events that may be viewed either from either the train
       or the embankment frames. There is therefore no conflict with Invariance of Contiguity. This point
        is further discussed in Section 6 below. In Einstein's interpretation it is assumed that only
        the events R'-D' (actually R-\DR\nolinebreak), G'-D'(actually  G-\DG)   and  R-G-D exist,
  and these events
      are considered only in the embankment frame.

       \par The following section presents an analysis of
        a more elaborate TETE that will be found to respect IC and which does indeed lead to the prediction of
        `relativity of simultaneity' of  certain events observed in different frames of reference, as a prediction
      of SR, but in a way quite different to that suggested by Einstein in the passage quoted above, or
      in the original SR paper. This analysis, in conjunction with the above critique of Einstein's 
       interpretation, will  also shed light on the correct operational meaning of the postulate E2, to be
       discussed in more detail in Section 5 below.

       \SECTION{\bf{The TETE as a photon clock: Two trains for relativity of simultaneity}}
        A version of the TETE approximating more closely an actual experiment will now be considered.
       A schematic is shown in Fig.4. There are two trains \TU and \TL  moving with velocities 
       $v$ and $w = c^2(\gamma_v-1)/(v \gamma_v)$ respectively in the positive x-direction relative to
         the embankment
        E. The latter velocity is chosen such that E and \TU have equal and opposite
        velocities, with absolute value $w$, relative to \TL. The rest frames of \TU, \TL and E
  are denoted by S', S'' and S respectively. 
        Both trains and the embankment are equipped with light sources G', R' Y'; G'', R'', Y'';
        \GU,  \RU, \YU;  \GL, \RL, \YL. The light sources  G', R'; G'', R'';
        \GU,  \RU;  \GL, \RL are connected to an electronic triggering system and sensors
        which fire the sources when  G', G'', \GU and  \GL are aligned with the point A
        on the embankment and  R', R'', \RU and  \RL  are aligned with the point B on the
        embankment. The length of the train is chosen so that all eight sources fire
        simultaneously in the frame S. This instant, which is the one corresponding to the 
        configuration shown in Fig.4, defines the origin of the time coordinates in all three
        frames of reference: $\tau = \tau' =\tau'' =0$. The photon detectors D', D'', \DU and \DL
        are situated mid-way between the corresponding G and R photon sources, and arranged in such
        a way that  D', D'', \DU, \DL detect only the photons emitted by the
        sources  \GU and  \RU,  \GL and  \RL, G' and R', G'' and R'', respectively.
         All of the detectors are  double-sided and equipped with coincidence electronics 
         such that, in the case that the photons are recorded, simultaneously, on both sides of the
         detector, the light sources  Y', Y'', \YU and  \YL will fire. The purpose of these
          sources is to enable the observation of coincidence events in one frame in both of the
          other frames. It will be found that it is precisely these coincidence events
          (not the G and R photon production events, that are simultaneous in all frames)
          that will show a `relativity of simultaneity' effect. As in the previous section, 
          the labels G, R, Y are used indiscriminately to indicate photons, or their sources.
           In the present TETE they can also be conveniently identified with the colours
           Green, Red and Yellow of the respective photons. Thus the Yellow signal indicates
           a space-time coincidence ($xt-$contiguity) of Green and Red photons. Suitable wavelength
         filters in front of the detectors can ensure that only appropriately coloured photons
         are detected.

 \begin{figure}[htbp]
\begin{center}\hspace*{-0.5cm}\mbox{
\epsfysize15.0cm\epsffile{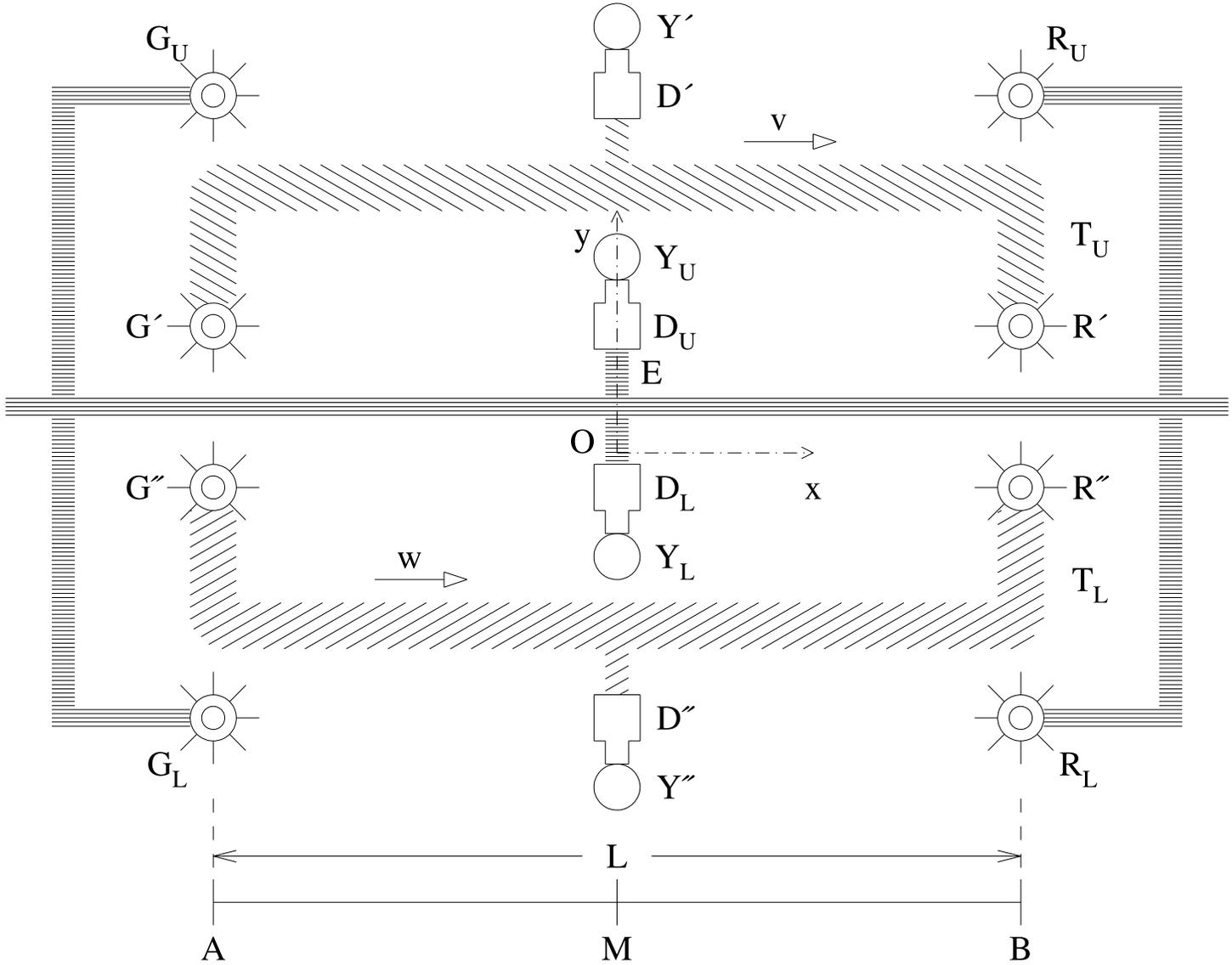}}
\caption{ {\em  A TETE with two trains \TU and \TL moving to the right with velocities $v$
   and $w = c^2(\gamma_v-1)/(v \gamma_v)$ respectively. The velocity $w$ is chosen so that,
    in the rest frame of \TL, \TU and the embankment E have equal and opposite velocities of
   absolute value $w$. Both trains and the embankment are equipped with light sources G and R and
  photon detectors D (see text for details). Each detector is also equipped with a light
   source Y that fires in the case that case that right- and left-moving photons are detected
   in coincidence.  The G and R light sources fire when they are all aligned with the points A and B
   on the embankment, as shown in the figure.}}
\label{fig-fig4}
\end{center}
\end{figure}

   \par The TETE shown in Fig.4 will now be analysed by applying Einstein's postulate E2 of SR
        in each of the frames S, S' and S''. In Fig.5 is shown the sequence of events in the train
        \TU (frame S') and on the embankment (frame S) as observed from the embankment, while Fig.6
        shows the sequence of events in the train \TU and on the embankment as observed from the train \TU.
        For example, at $\tau = 0$ (Fig.5a), the photons emitted from G' and R' start their transit towards
        \DU. Both of these sources are moving with velocity $v$ in the +ve $x$-direction relative
        to \DU. Similarly, in S' (Fig.6a) the sources \RU and \GU move in the -ve $x'$-direction
        relative to D' and the corresponding photons are emitted in the direction of D'. Application 
        of the postulate E2 in either frame predicts the observation of the Yellow signals shown
        in Fig.5b, 5c, 6b and 6c. At time $\tau = L/(2c)$ the photons G' and R' arrive in coincidence at
        \DU and the source \YU fires. The distance $L'$ between the sources
        G' and R' (compare with Fig.3a) can be found by use of the space-like\
       invariant interval $\Delta s$ 
         between arbitary events at A' and B' expressed in terms of space-time
         coordinate intervals in the frames S and S':
   \begin{equation}
   (\Delta s)^2 \equiv (\Delta x)^2-c^2(\Delta \tau)^2 =   (\Delta x')^2-c^2(\Delta t')^2
   \end{equation}
      In Fig.5a, $\tau = t' = \Delta \tau = \Delta t' = 0$, so from (4.1), 
    \begin{equation}
     \Delta s = \Delta x \equiv x_{B'}-x_{A'} = L =  \Delta x' \equiv x'_{B'}-x'_{A'} \equiv L'         
     \end{equation}
      This equation shows that the spatial separation of two physical objects at rest in
      an inertial franme is a Lorentz-invariant quantity~\cite{JHFPS} that takes the same value
      when measured in any inertial frame. There is no `relativistic length contraction' effect
      if the proper frame of the objects is in motion relative to another
      inertial frame~\cite{JHF1,JHF2,JHF3,JHF4}. 
     By considering events at A'' and B'' on the train \TL it is shown in a similar manner that
       $x''_{B''}-x''_{A''} \equiv L'' = L$. 
    It follows that \GU and G$_R$ will arrive at D' in S' at the time $\tau' = L/(2c)$. 
     Allowing for the TD effect, Eqn(2.11), (\TU constitutes a moving photon clock) the observer
     in S will see the coincidence of \GU and \RU at D' (i.e. the signal Y') at time
     $\tau = \gamma_v L/(2c)$ (Fig.5c). Thus, contrary to the conclusions of Einstein's analysis of
      a similar TETE,
      observers in both S and S' will judge the photon production events to be simultaneous.
     However, unlike for the case of sound signals, where terms of O($V_S/c)^2$) and higher are
    neglected, shown in Fig.2c, the \GU-\RU-D' coincidence is observed at a later time than 
   the G'-R'-\DU one. Note however, that the transit of the sound signals in the train in Fig.2
    constitutes a clock just as well as the photon transits in Fig.5, so that if terms
   of O($V_S/c)^2$) and higher are included the G'-R'-sound signal coincidence in Fig2c is also
    predicted to occur later than the G-R coincidence in this figure.

\begin{figure}[htbp]
\begin{center}\hspace*{-0.5cm}\mbox{
\epsfysize15.0cm\epsffile{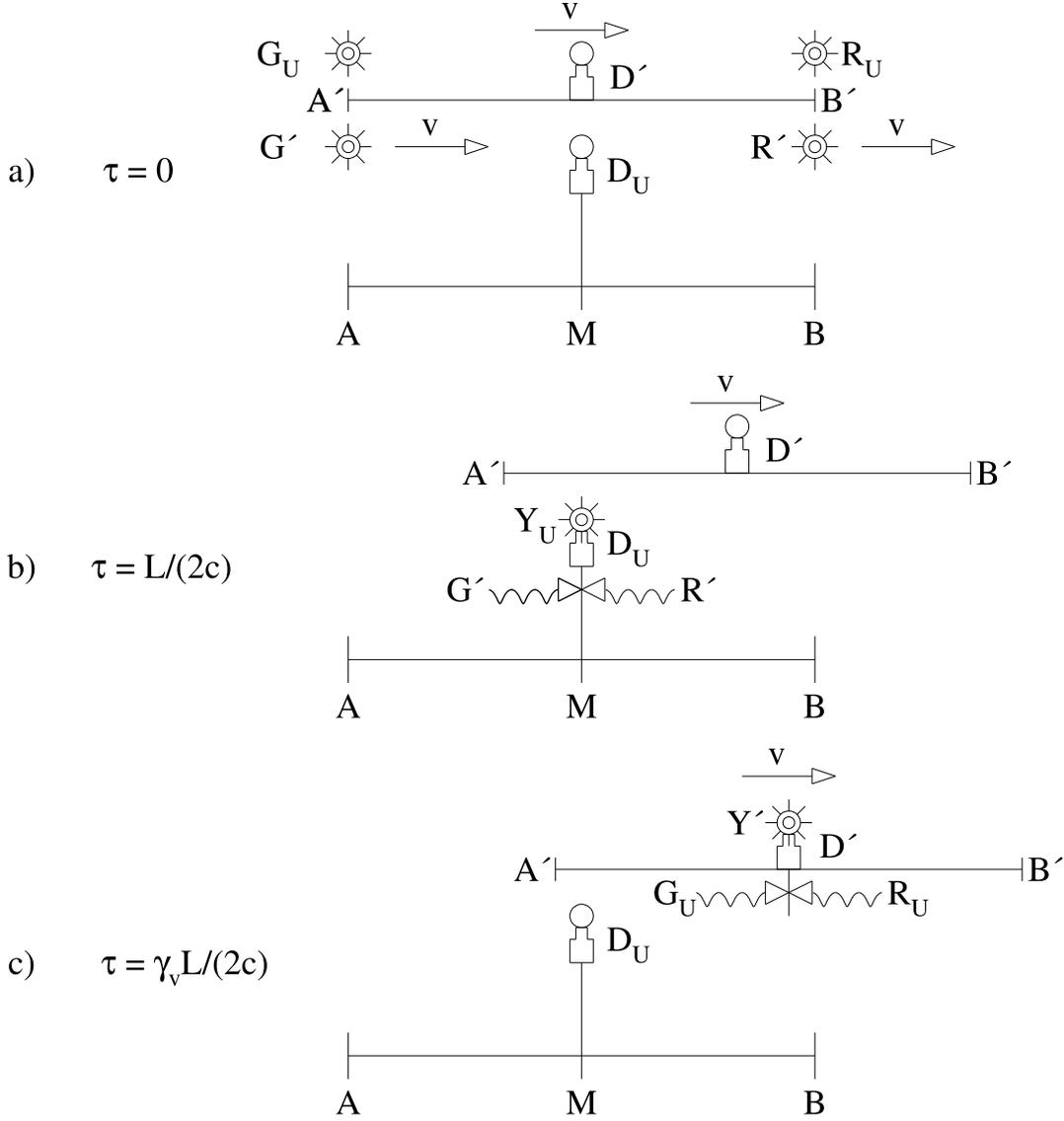}}
\caption{{\em  Events on the embankment E (frame S) and in the train \TU (frame S') as observed 
  from the embankment. a) Sources (G',\GU),(R',\RU) fire at $\tau = \tau'= 0$ when they are both
  aligned with  the points (A),(B) on the embankment. b) $\tau = L/(2c)$, photons from
  G' and R' are detected, in coincidence, by \DU; \YU fires. c) $\tau =\gamma_v L/(2c)$,
  photons from \GU and \RU are detected in coincidence by D' in S'; Y' fires and is observed from S.
  In this figure $\beta_v = 2/3$, $\gamma_v = 3/\sqrt{5}$. For clarity, the vertical positions of
   the detectors are shown displaced.}}
\label{fig-fig5}
\end{center}
\end{figure}

 \begin{figure}[htbp]
\begin{center}\hspace*{-0.5cm}\mbox{
\epsfysize15.0cm\epsffile{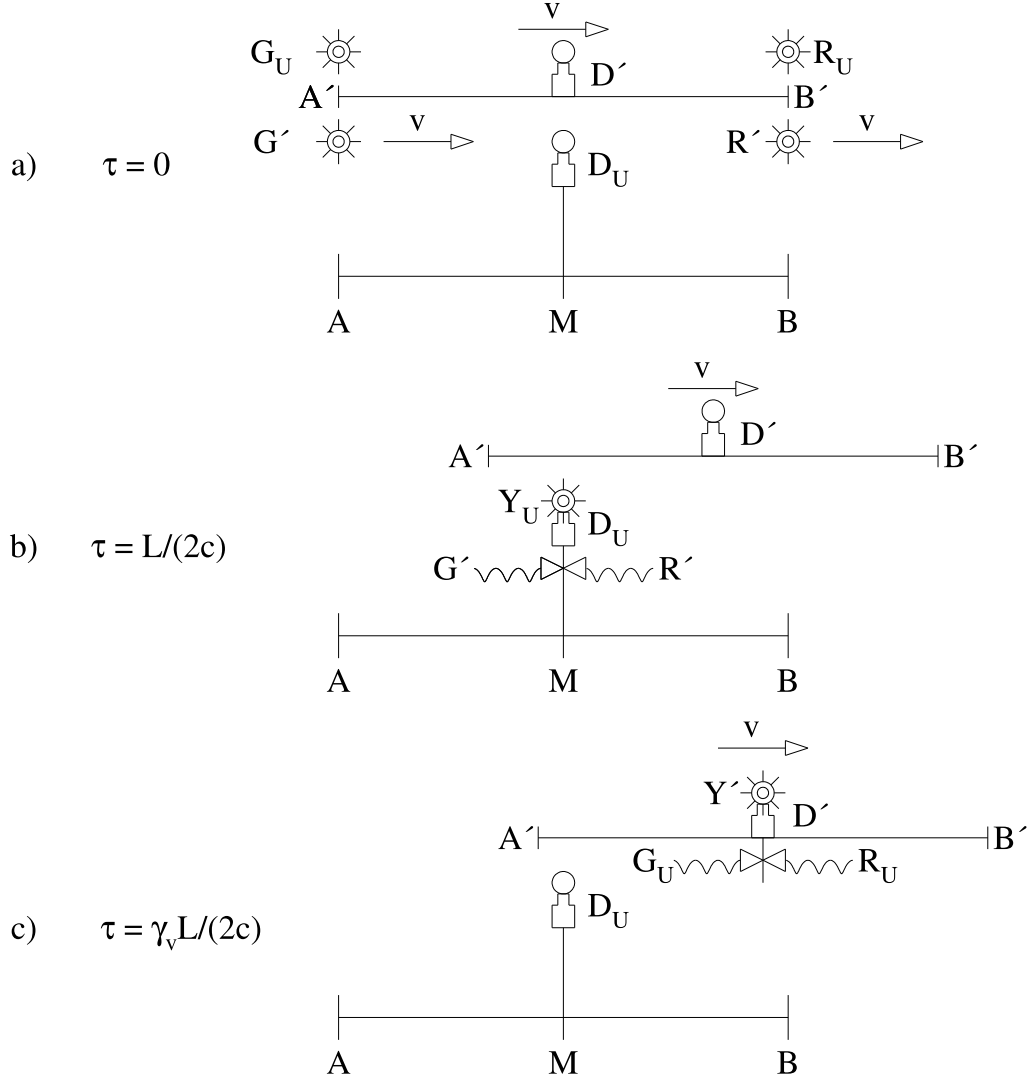}}
\caption{{\em Events on the embankment E (frame S) and in the train \TU (frame S') as observed 
  from the train \TU. a) Sources G', \GU, R' and \RU  fire at $\tau = \tau'= 0$. b) $\tau' = L/(2c)$,
 photons from \GU and \RU are detected, in coincidence, by D'; Y' fires. c) $\tau' =\gamma_v L/(2c)$,
 photons from G' and R' are detected in coincidence by \DU in S; \YU fires and is observed from S'.
 In this figure $\beta_v = 2/3$, $\gamma_v = 3/\sqrt{5}$. For clarity, the vertical positions of
   the detectors are shown displaced. }}
\label{fig-fig6}
\end{center}
\end{figure}
\begin{figure}[htbp]
\begin{center}\hspace*{-0.5cm}\mbox{
\epsfysize15.0cm\epsffile{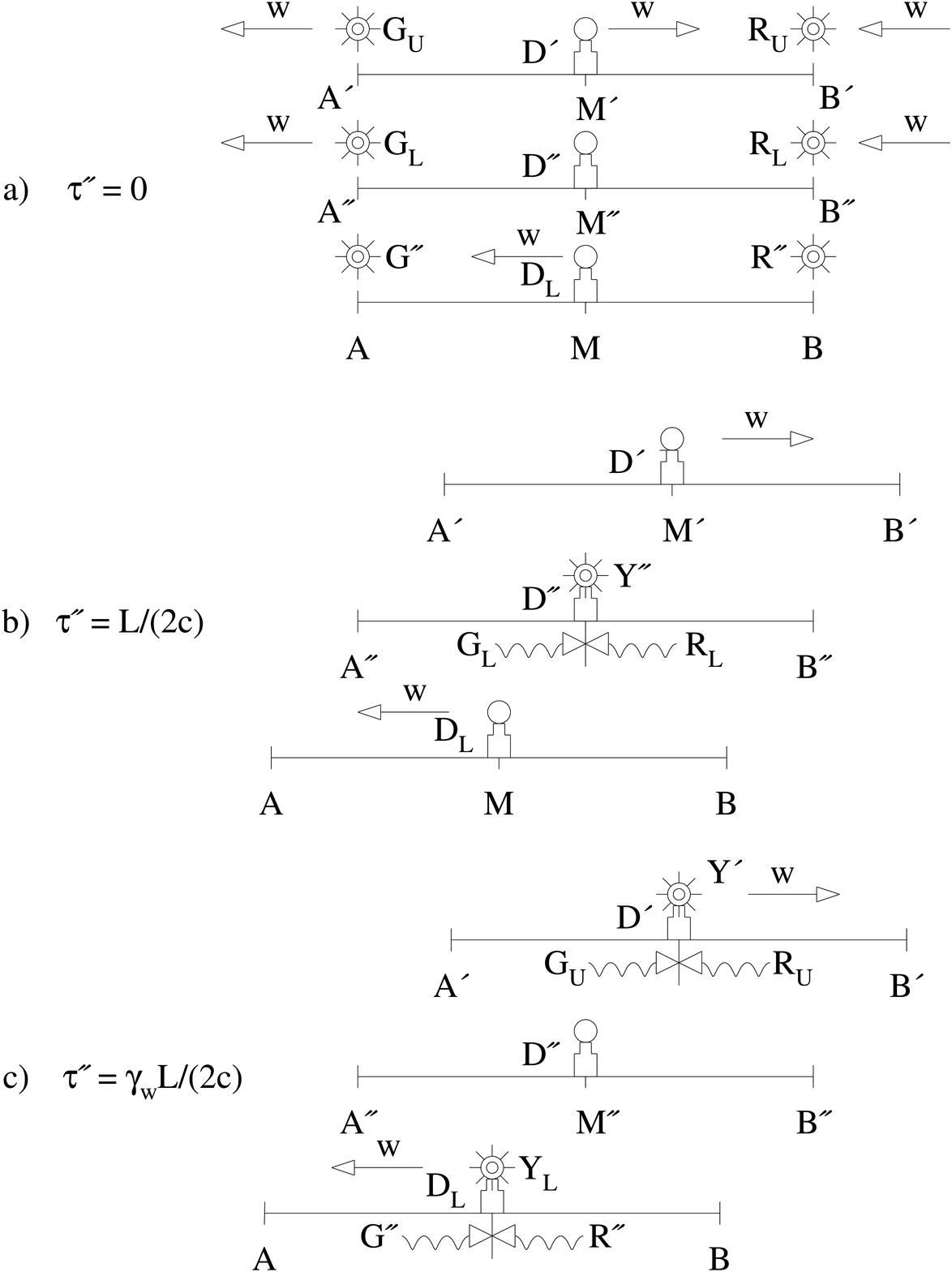}}
\caption{ {\em  Events on the embankment E (frame S), in the train \TU (frame S') and
  in the train \TL (frame S'') observed from the train \TL. a) Sources \GU, \GL , G'',
  \RU, \RL and R'' fire at $\tau = \tau'= \tau'' = 0$. b)  $\tau'' = L/(2c)$,  photons from
   \GL and \RL are detected, in coincidence, by D''; Y'' fires. c) $\tau'' = \gamma_w L/(2c)$,
   photons from \GU and \RU are detected in coincidence, by D'; Y' fires. Also photons from
   G'' and R'' are detected, in coincidence, by \DL; \YL fires. Y' and \YL are observed
   simultaneously in S''. In this figure $\beta_v = 2/3$, $\beta_w = 0.382$, $\gamma_w = 1.082$.
    For clarity, the vertical positions of the detectors are shown displaced. }}
\label{fig-fig7}
\end{center}
\end{figure}

       \par  Fig.6 shows the same sequence of events as in Fig.5, but as seen by an observer
        at rest in the train \TU. The  \GU-\RU-D' and G'-R'-\DU coincidences are observed
      as before, but now the former of these events is observed earlier, not later as in Fig.5. 
      The sequences of events in Figs. 5 and 6 are an example of an application of the MRP.
     One is obtained from the other by simple exchange of primed and unprimed quantities. Also, unlike
     in Figs.1 and 3, as interpreted by Einstein, IC is respected. 
     Observation of the signals Y' and \YU by observers
     on both S and S' shows that both of them agree that the R and G photons are detected 
     simultaeously, from which the simultaneity of the photon production events in both frames
     can be inferred.
    \par In Fig.7 is shown the sequence of events seen by an observer in the train \TL that moves
   with velocity $w$ along the +ve $x$-axis relative to the embankment and velocity $w$ along
   the -ve $x$-axis relative to the the train \TU. In Fig.7a the configuration of the sources 
   and detectors in the frame S'' at the time $\tau'' = 0$ is shown, together with their velocities relative to
   the  detector D''.  Using the photon transit
   times in the various frames as clock signals, and taking into account the TD effect (2.11) predicts
  the sequence of events shown in Fig.7b and 7c. The \GL-\RL-D'' coincidence
     (signal Y'') is observed at time  $\tau'' = L/(2c)$. and the  \GU-\RU-D' coincidence
      (signal Y') and the G''-R''-\DL coincidence (signal \YL) are observed simultaneously at the 
      later time  $\tau'' = \gamma_{w} L/(2c)$. Since it is clear, from the geometry of Fig.4, that the
      signals \YL and \YU  are $xt$ contiguous in S and therefore, by IC,  simultaneous in all frames,
     it follows that the signal events  Y' and \YU or \YL display relativity of simultaneity:
      \begin{eqnarray}
    {\rm Observed~from~E~(frame~S,~Fig.5)} &   & \tau({\rm Y'}) > \tau({\rm Y_U}) = \tau({\rm Y_L}) \nonumber \\
    {\rm Observed~from~T_L~(frame~S'',~Fig.7)} &
      & \tau''({\rm Y'}) = \tau''({\rm Y_L}) = \tau''({\rm Y_U}) \nonumber \\
  {\rm Observed~from~T_U~(frame~S',~Fig.6 )} &  & \tau'({\rm Y'}) < \tau'({\rm Y_U}) = \tau'({\rm Y_L}) \nonumber
    \end{eqnarray}
 Since the times of the signals Y', \YU, \YL and Y'' are found to be positive in all frames,
 and so later than the production time $\tau = \tau' = \tau'' = 0$ of the G and R photons, causality 
 is respected in all observations i.e. no observation of an apparent emission time that is later
 than an apparent detection time. 
 \par The above behaviour, showing relativity of simultaneity, derived from the synchronised
     `photon clocks' in S, S' and S'', will of course be displayed by any other, similarly synchronised,
      clocks in these frames. For example, clocks, of any construction, situated at the same
    positions as D', \DU, \DL and D'' and set to $\tau = \tau' = \tau'' = 0$ when they are all in
    $xt$-contiguity, as shown in the configuration of Fig.4.  
      
     \SECTION{\bf{Relative velocity and speed: The meaning of the second postulate}}
    The complete statement of the postulate E2 given in Ref.~\cite{Ein1} is: 
   \par {\tt Any ray of light moves in the ``stationary'' system of coordinates with the 
    determined velocity, c, whether the ray be emitted by a stationary or a moving body. Hence:}
     \begin{equation}
     {\tt velocity} = \frac{{\tt light~path}}{{\tt time~interval}}
     \end{equation}
     {\tt where `time~interval' is to be taken in the sense of the definition in
    Paragraph 1\footnote{That is, as defined by appropriate clocks synchronised by the
    light signal procedure defined in  Paragraph 1 of Ref.~\cite{Ein1}.}.} 
     \par This definition must be treated with some care and circumspection. In particular
     the {\it relative velocity} of two photons or of a photon and some massive physical
    object, as observed in a particular frame of reference, may well be less than, or
    greater than, $c$. 
     \par The definition of the mean relative velocity of of two objects A and B, over some interval
      of time $\Delta t = t_2-t_1$, as observed in some frame of reference, is:
      \begin{eqnarray}
 v_{BA} & \equiv & {\rm velocity~of~B~relative~to~A} \nonumber \\
        & \equiv & \frac{x_B(t_2)-x_A(t_2)-[x_B(t_1)-x_A(t_1)]}{t_2 - t_1}  \nonumber \\
       & \equiv  & \frac{\Delta x_{BA}(t_2)-\Delta x_{BA}(t_1)}{\Delta t}
    \end{eqnarray}
 where 1-dimensional motion is considered. For the case of three objects, each in uniform 
  motion along a straight line, the relation $\Delta x_{AB} = -\Delta x_{BA}$, following from
  the definition of $\Delta x_{AB}$, gives the identity relating three relative velocities:
   \begin{equation}
      v_{BA}+ v_{AC}+ v_{CA} = 0
   \end{equation}
   Consider the velocities of the photons R' and G' relative to the detector D' as viewed in the
   embankment frame S (Fig.1a). The geometry of the configurations shown in Fig.1a and Fig.1b give:
      \begin{eqnarray}
  v_{G' D'} & = & c-v \\
 v_{D' R'} & = & c+v \\
 v_{R' G' } & = & -2c
\end{eqnarray}
 The photons R' and G' certainly follow `light paths' in propagating from their points 
 of production on Fig.1a to their points of detection in Fig1b and Fig.1d respectively,
 however,in neither case is the appropriate velocity in Eqn(5.1) the constant $c$. 
  \par The relative velocities in Eqns(5.2)-(5.6)  ---that between two objects in
   the same reference frame--- must be distinguished from those appearing in the parallel
    velocity addition formula, which connects instead the velocities of a single physical
     object in different frames of reference. In each frame the velocity of the object
   as defined as that relative to the fixed coordinate axes of the frame. It is this velocity,
    and only this velocity, for the case of light, that is correctly given, in any inertial
    frame, by Einstein's formula (5.1). The speed of light, the magnitude of its velocity
     defined in this way, is then the same in all inertial frames. It is operationally defined
   as the distance between the fixed points of production and detection of a photon in
    an inertial frame, divided by the time-of-flight of the photon as measured by synchronous
    clocks at rest in the same reference frame.  
   \par It is interesting to note that Einstein did not explicitly state that the speed of 
      light measured in this way is the same in {\it all} inertial frames, when defining,
     in the passage quoted above, the postulate E2. This was done later in the paper during the
    derivation of the LT. Since Einstein regarded the constancy of the speed of light as
    a `law of nature', it was, for him, already implicit in the first postulate E1.
     \par In fact the postulate E2 is most easily understood not as a separate
     `law of nature', under which aspect it seems highly counter-intuitive
       \footnote{See Einstein's discussion of the parallel velocity addition
        formula in Chapters VI and VII of Ref~\cite{EinSGR}}, but rather as a simple
      kinematical consequence of the fact that light consists of massless particles. In
    the derivation of the LT in Ref.~\cite{JHFLT} in which electrodynamics or the propagation
    of light play no part, it is found that the relative velocity of two inertial frames, A and B,
    defined according to the formula (5.2) with either $x_A=$constant, or $x_B=$constant, has
    a maximum possible value, $V$. If one frame is the proper frame of an object of
    Newtonian mass, $m$, moving with velocity $v$ relative to the other, the  relativistic
   momentum, $p$, and energy, $E$, of the object in the second frame are defined by the relations:
        \begin{eqnarray}
       p & \equiv & \tilde{\gamma}_v m v \\
      E & \equiv & \tilde{\gamma}_v m V^2
   \end{eqnarray}
     where 
   \begin{equation}
   \tilde{\gamma}_v \equiv \frac{1}{\sqrt{1-(v/V)^2}}
   \end{equation}
    Eliminating successively $v$ and $E$ between Eqns(5.7) and (5.8) gives the relations:
       \begin{eqnarray}
      E^2 & = & m^2 V^4 + p^2 V^2 \\
    v & = & \frac{p V^2}{( m^2 V^4 + p^2 V^2)^{\frac{1}{2}}}
  \end{eqnarray}
   It follows from Eqn(5.11), that any massless object has velocity $v = V$ in any inertial frame.
   Identifying light with the massless photon then predicts that, in all inertial frames:
    \begin{equation}
  {\rm speed~of~light} \equiv c = V = {\rm constant}
   \end{equation}
   so that also $ \tilde{\gamma}_v = \gamma_v$ in all inertial frames. Eqn(5.12) is just Einstein's
   postulate E2 of special relativity.
   It follows directly from the identification of light with massless particles  --- for the
   discovery of which (not special relativity) Einstein was awarded the Nobel Prize.

    \par The definition (5.2) and the identity (5.3) can be used to discuss the apparent
     velocities, in the embankment frame S, of the photons \GU and G$_R$ that are detected
      by D' in the train \TU. Referring to Fig.5, the geometry of the configurations in Fig.5a
 and Fig.5c, Eqn(5.2) gives the relations:
    \begin{eqnarray}
     v_{G_U D'} & = & \frac{c}{\gamma_v} \\
  v_{R_U D'} & = & -\frac{c}{\gamma_v} \\
 v_{G_U R_U} & = &  \frac{2c}{\gamma_v}
   \end{eqnarray}
   Since $v_{D' D_U} = v$, Eqn(5.3) together with (5.13)
  and (5.14) can be used to derive the relations:
    \begin{eqnarray}
   v_{G_U D_U} & = & c\left( \beta + \frac{1}{\gamma_v}\right) \ge c \\
    v_{R_U D_U} & = & c\left( \beta -\frac{1}{\gamma_v} \right) \le c      
    \end{eqnarray}   
     The maximum value of $ v_{G_U D_U}$ or $\sqrt{2}c$ occurs when $\beta_v = 1/\gamma_v = 1/\sqrt{2}$.
     This is also the value of $\beta_v$ for which $ v_{R_U D_U}$ changes sign and the photon
       \RU apparently moves to the right, rather than the left in Fig.5. As $\beta \rightarrow 1$,
       $v_{G_U D_U} \simeq v_{R_U D_U} \rightarrow c$ so that the relative velocity of the
       photons \GU and \RU, as observed from S, vanishes.
       Comparison of the relative velocities (5.4) and (5.5) appropriate to Einstein's interpretation,
      uniquely in the frame S, of the TETE, with those (5.13) and (5.14) which describe instead
      the behaviour of the moving photon clock constituted by \TU, as observed from S,  provides
    the correct understanding in SR of both the `relativity of simultaneity' phenomenon 
     and the meaning of `velocity' in Eqn(5.1). The relativity of simultaneity effect describes
   different time ordering of the {\it same events} in different inertial frames, e.g. the 
    events Y' and \YU described in the previous section. In Einstein's interpretation
    of the TETE {\it different  events}, R'-D' (actually R-\DR), G'-D' (actually G-\DG)
    and G-R-D in the {\it same}
    inertial frame are compared. The $xt$-contiguity of  G-R-D  in S (Fig.1c) and the apparent
    lack of it in S' (fig.3b and Fig.3d) are at variance with IC, which is a necessary consequence
    of the LT. In Einstein's interpretation of the TETE both the photons which are detected in
     D (at rest in the frame S) and apparently detected in D' (at rest in the frame S') have velocities,
     relative to S,
    of $c$. However D' in this case is not a detector {\it in} S', it is instead identified with 
    detectors in S, \DR and \DG, that happen to be at the same spatial positions in S as D' when photons
    are detected. Indeed, the photons detected in D' cannot
     be the the same ones as those detected in \DR or \DG. When the postulate E2 is correctly invoked,
     as in the analysis
    presented in Fig.5, it can be seen that the apparent velocities in S of the photons
    detected in by D' in S', given by Eqns(5.13) and (5.14) are not, in general, equal to c,
    so that it is incorrect to calculate the times of coincidence of photons with D' by
     considering only photons with speed $c$ in the frame S.  This
     difference of apparent velocity is a consequence of the apparent slowing down of physical processes
      (in this case the speed of light progagation in an inertial frame) observed for all moving clocks
     as a consequence of the universal relativistic time dilatation effect\footnote{A similar discussion
    of the apparent velocity of photons in the arms of a Michelson interfermoeter, considered as a 
    `photon clock', may be found in Ref.~\cite{JHF1}.}. 

 \SECTION{\bf{Comments on some previous interpretations of the Einstein TETE}}
 A recent paper by Nelson~\cite{Nelson} presented an extensive re-interpretation of Einstein's 
  original TETE. The present author read this paper before writing the present one and found
   Nelson's analysis extremely useful in revealing the crucial questions which need to be 
   addressed in the problem. Nelson finally concluded that Einstein's interpretation was qualitatively correct,
   in that the lightning strokes appear to be simultaneous to the observer on the embankment
   and non-simultaneous to an observer on the train, but that actually working out the problem
   quantitatively, in the
    embankment frame, according to Einstein's prescription, gave a result for the time difference of the
   non-simultaneous events different from that found in Nelson's `relativistic' analysis, to be
   discussed below. This disagrees with the conclusion of the
 present paper that observers both on the embankment and in the train will judge that the strokes
   are simultaneous.

   \par Although Nelson's paper contains many important and true statements concerning the problem, 
       which will be discussed below, the reason for the difference, and why Nelson's (and Einstein's)
     conclusion is wrong becomes immediately obvious on performing a correct space-time analysis
     only of the lightning stroke events as observed on the embankment and in the train. Indeed, the
    the light signals introduced by Einstein are quite irrelevant to the question of the simultaneity
     (or not) of the lightning stroke events themselves\footnote{In Chapter VIII of
      Ref~\cite{EinSGR} Einstein mixes up the physical definition of simultaneous events with the problem of
      how a single observer would decide, experimentally, that two events were simultaneous. In connection
      with the latter, light signals were introduced. This discussion leads directly to the introduction
     of the TETE in the following chapter. Simultaneity can be conveniently
     defined by conceptually simpler methods e.g. by establishing a network of pre-synchronised clocks
     and comparing event times recorded by local clocks. It was the misunderstanding of the 
      physics of detection of the light signals, not of the lightning strokes, that lead Einstein
      to the erroneous conclusion of non-simultaneity of the train frame observations.}. 
         \par According to Einstein's original formulation of the TETE: (i) the lightning 
        strokes are simultaneous in the embankment frame. Einstein also states: (ii) `But the events
     A and B also correspond to the points A and B on the train'. Denote the lightning stroke
    events in the embankment frame as E$_A$ and E$_B$ and in the train frame as E$_{A'}$ and E$_{B'}$.
     The simultaneity of E$_{A'}$ and E$_{B'}$ as a consequence of that of E$_A$ and E$_B$ follows
    directly from the relation (2.11) connecting corresponding time intervals in S and S'.
     If $\Delta \tau_A = \Delta \tau_B$, according to synchronised clocks in S, 
     so that E$_{A}$ and E$_{B}$ and are simultaneous in S then also  $\Delta t'_A = \Delta t'_B$
     and  E$_{A'}$ and E$_{B'}$ are simultaneous in S'. The lightning strokes events 
    specified in the embankment frame are therefore
     simultaneous for observers both on the embankment and in the train as shown in Figs.1 and 3. This
     follows directly from  Einstein's statements (i) and (ii) that define the initial conditions of the TETE,
       and the TD relation
     (2.11). This conclusion is, of
     course, independent of any subsequent analysis of light signals produced by the lightning strokes.
  \par Because, unlike in the analysis presented above, using solely the coordinate-free TD relation
    (2.11), Nelson used the LT directly in his study of the TETE, the (coordinate-dependent)
     description of the experiment using the LT and synchronised clocks is now reviewed
     before coming to the critique of Nelson's arguments.  
      The essential point here is that the LT is, fundamentally, not a description of the
     `geometry of space-time' but of how similar clocks are observed to behave in different
      inertial frames \footnote{Just this remark was made by Einstein in Chapter X of
      Ref.\cite{EinSGR}.}. The relation
      \begin{equation}
         \tau(E_A) = \tau(E_B) = t'(E_{A'}) = t'(E_{B'}) = 0,
    \end{equation}
      as in the analgous TETE configuration shown in Fig.4, 
     corresponds to a particular choice of the synchronisation convention of clocks in the frames
     S and S'. Choosing the origins of the coordinate systems in S and S' at M and M', as done by Nelson,
      the synchronisation convention (6.1) corresponds to the following space-time transformations
     for clocks situated at A', M' and B' respectively:
    \begin{eqnarray}
     x'_{A'}+L/2 = \gamma_v[ x_{A'}+L/2- v \tau_{A}] = 0 \\
       t'_{A'} =  \gamma_v[ \tau_{A}-\frac{v( x_{A}+L/2)}{c^2}] \\
     x'_{M'} = \gamma_v[ x_{M'}- v \tau_{M}] = 0 \\
  t'_{M'} =  \gamma_v[ \tau_{M}-\frac{v x_{M'}}{c^2}] \\
  x'_{B'}-L/2 = \gamma_v[ x_{B'}-L/2- v \tau_{B}] = 0 \\
       t'_{B'} =  \gamma_v[ \tau_{B}-\frac{v( x_{B'}-L/2)}{c^2}]
  \end{eqnarray}
    Choosing $\tau_{A} = \tau_{M} = \tau_{B} \equiv \tau = 0$ when $x_{A'} = -L/2$,  $x_{M'} = 0$
       and  $x_{B'} = L/2$, as in the configuation shown in Fig.4, gives
   $t'_{A'} = t'_{M'} = t'_{B'} \equiv t' = 0$, which is just
 the synchronisation convention (6.1) above. With this convention Eqns(6.2) to (6.7) simplify
 at S time\footnote{This is the time registered by all synchronised clocks in S.}$\tau$
   to :
  \begin {eqnarray}
 x'_{A'} & = & -L/2,~~~  x_{A'} =  -L/2 + v\tau,~~~t'_{A'} = \frac{\tau}{\gamma_v} \\
  x'_{M'} & = & 0,~~~  x_{M'} =  v \tau,~~~t'_{M'} = \frac{\tau}{\gamma_v} \\
  x'_{B'} & = & L/2,~~~  x_{B'} =  L/2 + v\tau,~~~t'_{B'} = \frac{\tau}{\gamma_v}
  \end{eqnarray}
   Thus 
\begin{equation}
  t'_{A'} = t'_{M'} = t'_{B'} = t' = \frac{\tau}{\gamma_v} 
 \end{equation}
  no `relativity of simultaneity' 
 and 
 \begin{equation}
   x'_{B'}- x'_{A'} =  x_{B'}- x_{A'} = L
 \end{equation}
   no `length contraction'.
  The relations (6.11) and (6.12) were previously derived above (Eqn(2.11) and (4.2) respectively),
   without explicit use of the LT, by considering invariant space-time intervals in the frames
   S and S'.
    \par After this preliminary discussion a survey of the arguments presented in Ref.~\cite{Nelson}
     is made. The crucial statement in the abstract:
     \par {\tt It is shown that, under the conditions of the experiment, a simultaneity in one
     inertial frame does indeed hold in another inertial frame, but it is preceived as a 
      non-simultaneity by an observer outside the given frame.}
     \par is true and important for the understanding of the problem addressed, but
   its meaning depends crucially on that assigned to the word `perceived'. I return to this 
   point below.
   \par I agree with almost everything in Nelson's Sections 2-4. Indeed, since Einstein's
    analysis is carried out only in the embankment frame (there are no primed coordinates or
    times) nothing can be said concerning observations in the train frame. If the corresponding
    events are not even considered in the analysis it is evidently not possible to discuss
    simultaneity (or lack of it) in the train frame S'. To put it bluntly, Einstein's analysis
     does not even address the question posed by the TETE. For this, coordinates and times in
     the train frame must be considered. This was clearly and correctly stated by Nelson
     in his Section 3. In fact the `non-simultaneity' assigned by Einstein
    to the observations of the train observer has a trivial explanation ---the different
    relative velocities of D' relative to the photons G' and R' in the frame S (see Fig.1)
    \footnote{In discussing points raised in Ref.~\cite{Nelson} the notation of the present
     paper is used}, so, as explained above, what is shown in Fig.1 is the interaction of photons in the frame
     S (i.e. moving with the speed $c$ in this frame, according to the postulate E2) with
     {\it detectors at rest in S at the same position as the
     moving detector D' in this frame at the instant of photon detection}.
      It tells nothing of the response of D' to photons
     that are moving, according to E2, with the speed $c$ in the frame S'. This is what is shown 
    in Fig.3. Here D is in motion in S' with different velocities relative to the photons G and R
     and the `non-simultaneity' of its apparent\footnote{As in S, the corresponding S' frame
    events are actually specified, as discussed Section 3 above, by the concidences of photons
    with detectors at rest in S'
    at the same positions as the moving detector D at the S' times of the coincidences.} 
    encounters with them, is the same purely classical effect
     \footnote{It is classical because precisely the same result is obtained in Galilean relativity,
       where photons are replaced by massive objects of constant and equal velocities. The same pattern
     of events is seen in Fig.2 where light signals are replaced by sound signals. The underlying
     physical description of sound propagation is also purely classical.},
     following from the definition of relative velocity in Eqn(5.2), as seen in Fig.1. 
      \par Nelson's digression 4.1 is irrelevant to the discussion since, as shown above,  
     the question of simultaneity or non-simultaneity of the lightning strokes in the frames
     S and S' (before any consideration of light signals) has the same answer ---yes they are
     simultaneous in both frames--- whether a Galilean or Lorentz transformation is used. The simple reason
    is that the effect of the LT must be the same at the points A and B, 
    --- this is just translational invariance~\cite{JHF1}.
    So if $\tau_A = \tau_B$ then necessarily $t'_A = t'_B$. If the same synchronisation convention
    \footnote{ That is, the initial settings $\tau_0$ and $t'_0$, as viewed from S, of clocks at rest in
   S and S' respectively.} is used at
    A and B this will always be the case. Thus the statement in Section 4 of  Ref.~\cite{Nelson}:
    \par {\tt Put another way A(embankment) and A(train) are space-time coincident  ---i.e. they are
    one and the same point--- when the strokes occur, so that a stroke that hits  A(embankment)
   also hits  A(train) at the same instant. Likewise the stroke hitting  B(embankment) also hits
    B(train).}
   \par is true in both Galilean and special relativity, on twice replacing `also hits' by `is also observed from'.
   It is essential only that the strokes `hit' the embankment, not also the train.
    This was the way Einstein's TETE was defined in Ref`\cite{EinSGR}. Indeed (see the Appendix) depending 
   of how precisely the `strokes' are defined, they may not even hit the ends of the train at any time.
   Also, as shown above, `the train spans the same
  length in each frame' --- there is no length contraction. Thus Nelson's remarks in Section 4.1 on the 
 `pre-relativistic'
   nature of the analysis presented later in Section 7 of Ref.~\cite{Nelson} are ill-founded. Nelson's
  Figs.5 and 6 describe correctly the problem in both Galilean and special relativity. Only the 
  physical meaning of the photon-detector coincidence events (in the language of the present
  paper) is misinterpreted by Nelson.
   
    \par I find the `Traditional' analysis of Section 6 of Ref.~\cite{Nelson} to be quite  spurious.
     Non-simultaneous events observed in the frame S due to different photon-detector relative
    velocities are transformed into the frame S' and interpreted in terms of photons
    observed in this frame (i.e. photons with speed $c$ in S'). I can see no physical rationale for such
     a procedure.
     \par The analysis presented in Sections 7 and 8 of Ref.~\cite{Nelson} is very similar
      to that in Section 3 of the present paper. Figures 5 and 6 of  Ref.~\cite{Nelson}
      are essentially the same as Figs.1 and 3 of the present paper, the same photon-detector
      coincidence times being found. In Section 7 is found the statement: `This would seem to show that 
     simultaneity in one frame does indeed hold in another frame, but does not appear this way to
     an observer outside the frame'. This point is the same as that quoted above from the abstract
    of Ref.~\cite{Nelson}. The photons G and R  are detected simultaneously in S by D, whereas the
    other photons G' and R' apparently encounter the detector D' at different times in S (Fig.1)
    On the other hand
     the photons  G' and R' are detected simultaneously by D' in S' whereas the photons G and R
    apparently encounter the 
     detector D at different times in S' (Fig.3). The phrase `does not appear this way to
     an observer outside the frame' refers, for example, to the S' frame observation of temporally distinct
    G-D and  R-D coincidences instead of the triple G-R-D coincidence observed in S. But of course, as explained
     above
    {\it these are not the same events}. The photons G and R detected in S are not the photons
       G and R detected in S'. Also the  moving detector D in S' is actually strictly equivalent to detectors {\it
      at rest in this frame} which happen to be at the same spatial position as the detector D 
      at the instant at which the photons are detected. Only in this case can a meaningful time (that in S)
       be assigned to the corresponding events.  
     \par The pattern of events shown in Figs.1 and 3 is, as previously stated, consistent 
        with the Measurement Reciprocity Postulate. Contrary to what is stated in Section 7 of
    Ref.~\cite{Nelson} there is no paradox to be resolved here. What are shown in Table 1 of
     Ref.~\cite{Nelson} are not inconsistent observations of the lightning strokes but the 
     predicted and compatible time differences  of {\it different} photon-detector coincidence events
     \par In Section 9.1 of Ref.~\cite{Nelson}  the important  remark is made that space-time
       coincidences of different
    events are invariant and so their existence must be agreed upon by observers in any frame.
    --- this is just the `Invariance of Contiguity' proved in Section 2 above. It is suggested
      (as is done in the refined TETE of Section 4 of the present paper) to generate a third light
     signal if a double light signal-detector coincidence is recorded (for example the G'-R'-D'
      coincidence of Fig.3 of the present paper). Of course, as is shown in Fig.5 (which is analogous
      to Fig.1 for the original TETE) the embankment observer {\it does} see the signal Y' that
      shows that the analogous \GU-\RU-D' coincidence does occur in the train. Similarly, as shown in Fig.6,
      the train frame observer sees the \YU signal indicating the G'-R'-\DU coincidence on the embankment
     which  is analogous to the G-R-D one of Fig.1. Contrary to what is said in Section 9.1 of
       Ref.~\cite{Nelson}
      there is no `repudiation' of invariants. The mistake here is the identification of the separate
      R'-D' and G'-D' coincidence events of Fig.1b and 1d respectively in the frame S with the triple
      coincidence event  G'-R'-D' of Fig.3c in the frame S'. The photons R' and G' of Fig.1, that
      are detected in S are not the same as R' and G' in Fig.3, observed in coincidence in the frame S',
       The three coincidence events R'-D'-not G' (Actually R-\DR), G'-D'-not R' (Actually G-\DG) in S and
       G'-R'-D' in S' all occur,
       {\it but they are all distinct events}, not the same events as viewed in different frames, as assumed
      by Nelson, so that Invariance of Contiguity is respected.. 
      \par Contrary then to what is stated in Section 10 of Ref~\cite{Nelson} the patterns  of
     events shown in Figs.5 and 6 of Ref~\cite{Nelson} or in Figs.1 and 3 of the present paper are perfectly
     consistent. There are three distinct events for observers in either the embankment or the train frames:
     two single photon-detector coincidences and one double photon-detector coincidence, giving, in total,
      {\it six different events} seen directly in the embankment and train frames. The fundamental error of
     both Einstein and Nelson is the attempt to analyse the problem in terms of only three events specified only
     in the embankment frame. The mistake made  here is similar to that, mentioned in Section 2 above, where it
     is attempted to analyse the TD effects for observers in the frames S and S' uniquely in terms of $\Delta \tau$
      and $\Delta \tau'$ instead of the four physically distinct time intervals $\Delta \tau$, $\Delta t$, 
     $\Delta \tau'$ and $\Delta t'$.
      \par The three coincidence events observed in each of the embankment and train frames, may, if light
       signals are also generated by the single photon-detector coincidences (as described in Section 5 above
      for the double
      coincidence events) be observed from the other frames. This gives, in each frame, six events:
      the three coincidences recorded by the detectors at rest in the frame, and three signals
     produced by the similar coincidence events in the other frame ---in all twelve distinct
     events instead of the three assumed by Einstein and Nelson. It should be obvious that all these
      events are distinct. Six correspond to `baking a cake', three on the embankment and three
      on the train while six correspond to `watching the cake bake', again three on the embankment and three
      on the train. Of course the cakes being baked on the embankment and in the train are
      quite distinct. The times of all twelve events are presented in  Table 1.

   \begin{table}
\begin{center}
\begin{tabular}{|c||c|c|c|c|c|c|} \hline
 Event   & R-\DR & G-R-D & G-\DG & G'-\DG' & G'-R'-D'&  R'-\DR'  \\
 \hline 
  Time in S, $\tau$ &  $\frac{L}{2(c+v)}$ & $\frac{L}{2c}$ & $\frac{L}{2(c-v)}$ & $\frac{\gamma_v L}{2(c+v)}$
   & $\frac{\gamma_v L}{2c}$ & $ \frac{\gamma_v L}{2(c-v)}$    \\
 \hline 
  Time in S', $\tau'$ & $\frac{\gamma_v L}{2(c+v)}$ &  $\frac{\gamma_v L}{2c}$ & $ \frac{\gamma_v L}{2(c-v)}$ &
 $\frac{L}{2(c+v)}$ & $\frac{L}{2c}$ & $\frac{L}{2(c-v)}$ \\
 \hline
\end{tabular}
\caption[]{{\em Observation times on the embankment (frame S) and in the train (frame S') of the six distinct
       photon-detector coincidence events.}}      
\end{center}
\end{table}
  \par In Nelson's `relativistic analysis' presented in Section 11 and the appendix of Ref~\cite{Nelson},
   as in Einstein's analysis, only events defined in the embankment frame are considered, but unlike
   Einstein, the LT is used to transform these events into the train frame. In the notation
   of the present paper, the events R-\DR and G-\DG are identified with events R'-D' and G'-R', respectively
   in the train frame and it is concluded, by Lorentz transforming these events into the train
    frame, that there is no R'-G'-D' coincidence event in the latter frame. A complicated calculation
     is then performed transforming the lightning stroke events from S into S', as well as finding the
    propagation delays between light signals in S' produced by the transformed 
    stroke events and the moving detector D in this frame, to
    show that the double coincidence event R-G-D in S is also observed as such by the train observer.
     Actually, this follows directly from the application of IC to this event without any consideration
   of the prior lightning strokes and photon propagation. The spurious `relativity of simultaneity'
  effect given by incorrect use of the LT (6.4) and (6.5) appropriate to M and M' to transform the lightning 
  stroke events, (instead of (6.2) and (6.3) for the stroke at A and (6.6) and (6.7) for the stroke at B),
  predicts that the lightning strokes will be seen with a time diffrence
   $\Delta \tau'_{RS} = \gamma_v v L/c^2$ by the train observer. Nelson claims that this time difference
   is equal to that between the R-\DR and G-\DG coincidences (wrongly identified with R'-D' and G'-D'
    coincidences observed by the train observer). Taking into account the TD effect of Eqn(2.12) the train 
   observer will see, from the entries of Table 1, a time difference between these events of:
   \begin{equation} 
     \Delta \tau'_{TD} = \frac{\gamma_v L}{2}\left[ \frac{1}{c-v}- \frac{1}{c+v}\right] =\frac{\gamma_v^3 L v}{c^2}
    \end{equation} 
    In order to get agreement between $\Delta \tau'_{RS}$ and $ \Delta \tau'_{TD}$ Nelson applied the TD
   effect in the opposite direction, i.e. the formula $\Delta t = \gamma_v \Delta \tau'$ was used instead 
    of the correct formula (2.12). Thus, by dint of not considering events in the train frame, invoking
    the spurious `relativity of simultaneity' effect and an incorrect calculation of TD, Nelson arrived
   at a result in qualitative agreement with Einstein's one.
    \par To summarise Nelson's paper: It was correctly pointed out that Einstein's interpretation,
     considering events only in the embankment frame says nothing about what an observer on the train will 
     see, as events in the train frame are not even considered. Assuming simultaneous lightning 
     strokes on the embankment and in the train, and no length contraction (stated by Nelson 
    to correspond to a `pre-relativistic' analysis, but actually the correct 
     prediction of both Galilean and special relativity) the similar sequences of events on the embankment and 
     in the train are then considered (Section 7 and Figs. 5 and 6 of  Ref.~\cite{Nelson}). From these it
    is correctly concluded that both embankment and train obervers see simultaneous events. It is then
     assumed (in the notation of the present paper) that the photon-detector coincidences R-\DR, (called R'-D')
     and G-\DG  (called G'-D') and R'-G'-D' are different observations of the same event:  R'-G'-D' in S'
      and  R'-D' and G'-D' in S. This is in clearly in contradiction with Invariance of Contiguity, and so
      Nelson concludes that the  R'-G'-D' coincidence in S' does not exist. In view of the
     reciprocal nature of the events shown in Figs.5 and 6 of Ref.~\cite{Nelson}, it could have been
     argued, by the same logic, that the R-G-D coincidence in S does not exist. The fallacy of such arguments
     is, for example, the incorrect identification of G-\DG with G'-D' and  R-\DR with R'-D'. Actually  G-\DG and  R-\DR
      in S and R'-G'-D' in S' are distinct and uncorrelated events (different photons are detected in 
      each case). The `relativistic' solution suggested has been discussed above. Only events in the embankment
     frame are considered, the spurious `relativity of simultaneity' effect, resulting from misuse of the LT, is
     invoked, and the TD effect is incorrectly calculated, to yield an interpretation in qualitative
     agreement with Einstein's one.
      \par Nelson's correct insistence on the necessity of analysing the problem in both the train and 
      embankment frames was contested in Comments by Rowland~\cite{Rowland} and Mallinckrodt~\cite{Mallin}
      Unlike Nelson, no attempt at a quantitative analysis of the problem was proposed, so that only verbal
      arguments were presented. Rowland correctly stresses the importance of Invariance of Contiguity for
      the problem but wrongly concludes from this that embankment frame analysis only, as in Einstein's
        interpretation, is sufficient. Only embankment frame events are discussed, so the essential problem
       in not addressed. Similarly, Mallinckrodt considers only events in the embankment
      frame but makes the additional and incorrect statement (rebutted in Nelson's reply~\cite{NREP} to the
      Comments) that the postulate E2 is essential for Einstein's interpretation. None of Nelson, Rowland and
      Mallinckrodt understand the essential point that {\it different events} (i.e. events initiated by
      different photons) are observed in the embankment and train frames, so that it fallacious to assume
       that they are the same events viewed in different frames. This fact is a consequence of relativistic
       kinematics, one important aspect of which is encapsulated in Einstein's postulate E2. If the photons
       in the TETE
      were replaced by massive particles obeying Galilean kinematics Einstein's interpretation would
      be correct and the three events R-\DR, G-\DG and G-R-D would be observed at the same time in both
      the train and embankment frames and would constitute a complete and correct description of the
      problem.  But one would not conclude from this, any more than for the sound signals in Fig.2, 
      that `relativity of simultaneity' an essential feature of Galilean Relativity or of acoustics.
      \par The essential point, that it is not possible, in special relativity, to associate the times
      of coincidence events in detectors at rest in a reference frame with the proper time of an event
      in a spatially contigous but moving detector is made particularly clear in the example discussed
      below in the Appendix in which the `lightning strokes' of Einstein's TETE are replaced by two pulsed 
      laser beams in the embankment frame. 

       \SECTION{\bf{Summary and conclusions}}
 In Einstein's TETE the lightning strokes at A and B (see Fig.1a) are, by definition, 
  simultaneous in the embankment frame. It then follows directly from the TD formula (2.11)
 that these events are simultaneous in the train frame, i.e. if $\Delta \tau_A = \Delta \tau_B$
  in (2.11) then necessarily $\Delta t'_A = \Delta t'_B$. This is exemplified by comparison
  of Fig.8b and Fig.12 in the Appendix ---the events corresponding to the photon bunches striking the points A and
  B are simultaneous in both the embankment and the train frames. However, in order to analyse
   the problem, Einstein did not consider directly the lightning stroke events but rather light
   signals produced by the strokes and later seen by observers at the middle of the train or at
   point on the embankment midway between A and B. Considering only events in the embankment
    frame, as shown in Fig.1, it was concluded that the embankment observer
   would see the signals from the lightning strokes simultaneously (Fig.1c) but not
    the train observer (Fig.1b and Fig.1d). The latter then infers that the 
   lightning strokes must occur at different times in the train frame. Einstein took this
   inference of the supposed observations of the train observer as a statement of
    fact about any two events in an inertial frame~\cite{EinSGR}:
    \par {\tt Events which are simultaneous with reference to the embankment are not
    \newline simultaneous with respect to the train and vice versa (relativity of
     \newline simultaneity).}
     \par This interpretation is based on the assumption that the times of events
    in the train frame may be identified with the times of events observed in the embankment
    frame at the same apparent spatial positions. For example, in Fig.1b, the time of the train frame
    coincidence event R'-D' is assumed to be the same as the embankment frame event R-\DR.  As shown by
   the calculation presented in the Appendix, this would  be true if the photons in Fig.1 would
   be replaced by massive particles obeying Newtonian kinematics. However this assumption is false
   (and therefore invalidates Einstein's argument) for massive or massless particles obeying relativistic
    kinematics. Fundamentally, it is this neglect of relativistic kimematics, embodied, for photons, by
    the postulate E2, that leads Einstein to a conclusion that is the opposite of the correct one.
     Using E2, and performing the analysis in the train frame, as shown in Fig.3, leads, in combination
     with the embankment frame analysis of Fig.1, to the 
    correct conclusion that the lightning strokes are observed to be simultaneous by both the
    embankment and the train observers.
     \par   As shown by the analysis of the TETE with two trains presented in Section 4, a genuine `relativity
     of simultaneity' effect does exist, in special relativity, for certain events, in {\it different}
       inertial frames.
     However, as shown above, it follows directly from the TD effect formula that two events which are 
     simultaneous in the {\it same }inertial frame are so in all such frames ---contrary to Einstein's statement
     quoted above.
     \par In Section 5, the precise operational meaning of the postulate E2 is discussed and the important
    distinction between `speed' and `relative velocity' is pointed out. It is also shown that E2 does not 
    need to be introduced as a separate (and anti-intuitive) hypothesis, in special relativity
    but is direct consequence
    of the relativistic kinematics of massless photons~\cite{JHFLT}. Derivations of the LT
    not using the postulate E2 are briefly mentioned in Section 2 where the TD effect and
    Invariance of Contiguity are introduced  Due to the TD effect, the apparent speed of
     light signals between events, observed in a moving inertial frame from another one,
     may be greater than, or less than, the light speed, $c$,
     as measured in an inertial frame (Eqns(5.16) and (5.17)). 
    \par Section 6 contains a detailed discussion of Nelson's~\cite{Nelson} re-interpretation of Einstein's TETE.
       Nelson correctly pointed out the impossiblity to analyse the problem in a meaningful manner by considering
      only events
      in the
      embankment frame. The correct solution of the problem was presented in Nelson's Section 7 and shown in
      his Figs.5 and 6 (the latter being equivalent to Figs.1 and 3 of the present paper). However, this solution
      was rejected by the failure to notice the distinct nature of (in the language of the present paper)
      the photon-detector coincidence events in the different frames. See Table 1 and the accompanying
       discussion in the text. These were misinterpreted by Nelson as observations, in different frames, of
        the same events, with a resulting apparent `repudiation of invariants' or violation of
         Invariance of Contiguity. This lead to the rejection of the correct interpretation shown in 
         Figs.5 and 6 of Ref.~\cite{Nelson} and the introduction of the `relativistic analysis' of Nelson's
         Section 11 and Appendix. This, like Einstein's interpretation, considered only coincidence events
         in the embankment frame, which were, however, subsequently Lorentz-transformed into the train frame.
          In the calculation the spurious `relativity of simultaneity' effect (correlated with the equally
          spurious `relativistic length contraction' effect) resulting from misuse of the space-time
           LT~\cite{JHF1,JHF2,JHF3,JHF4} was invoked, and the TD effect of Eqn(2.12) between the train and 
           embankment frames was inverted. This incorrect calculation gave a result in qualitative
         agreement with Einstein's interpretation.
         \par In the Appendix of the present paper, instead of considering observations of light signals 
          produced by the `lightning strokes' of Einstein's TETE, the latter are replaced by temporally
         localised bunches of laser photons generated in the embankment frame (see Fig.8a). Applying Einstein's
         interpretation to the configuration of Fig.8b, where the bunches of photons hit the points A and B
         on the embankment, it would be concluded that the former hit simultaneously the $xt$-contiguous points 
          A' and B' on the train. The detailed space-time analysis of the problem presented shows that this would
          indeed be the case if the photons were replaced by massive particles moving at a constant velocity
          $u$, $\ll c$, but this is no longer the case when $u$ is of the order of $c$. Comparison of Figs.10.
          11 and 12 shows the profoundly non-intuitive nature of the spatial configurations of moving objects 
          at corresponding times (i.e. times linked by the TD formulae (2.11) or (2.12)), in different
           frames, as predicted by special relativity theory.
   \par {\bf Added Note}
   \par The conclusion of this paper, contrary to Einstein's assertion, that observers both on
     the embankment and in the train will find that the lightning strikes on the embankment occur
      simultaneously is a valid one, but the arguments given to reach this conclusion are flawed
      by a fundamental conceptual error. Indeed, all the necessary information to obtain the correct 
      solution is presented in Sections 2 and 5 together with the relation $L = L'$ that is, however, (see below)
      incorrectly derived in (4.2). 
      \par The error which is made (in common with all previous analyses of the problem, including that of
      Ref.~\cite{Nelson})
       is the conflation of kinematical configurations in a primary experiment and its reciprocal, related  by the MRP 
       as discussed in Section 2, with observations in the frames S or S' of either one of these experiments which,
       as correctly stated in Section 2, are {\it physically independent}. In the primary experiment, clocks at rest
       in S' are seen to run slower than clocks in S (Eqn(2.11)) whereas in the reciprocal experiment, clocks at rest
       in S are seen to run slower than clocks in S' (Eqn(2.12)). As pointed out in connection with Eqns(2.13) and (2.14)
        it is impossible that clocks can run both fast and slow in the same experiment. 
        \par Since Einstein described the observations of coincidence events (between the world lines of 
        observers and light signals) only in the embankment frame, these are the events that must be transformed into the
        train frame in order to analyse the experiment. Einstein did not do this and, as inspection of Figs. 5, 6 and 7 
       shows, neither did the present author! What are correctly shown in these figures are instead configurations
       of physically-independent primary and reciprocal longitudinal photon-clock experiments as previously
        discussed in connection with the Michelson Interferometer in Ref.~\cite{JHF1}. The same mistake, arising
        from misinterpetation of the postulate E2, occurs in Fig.~4 of  Ref.~\cite{Nelson}. If the observed speed of light
        actually were $c$, independently of the speed of its source, in all inertial frames, as naively follows from the
        postulate E2, then the description of the signals G' and R' shown in Fig. 5 is correct. However 
       the signals \GU and \RU which, by definition, have speed $c$ in the frame S' have, as discussed in Section 5,
       the observed speeds in S: $c^{\pm}_{app} = c (\beta \pm 1/\gamma_v)$) (Eqns(5.16) and (5.17)).  As the
       Relativistic Relative Velocity Transformation Relation (RRVTR) of which these formulae are examples, is a necessary
       consequence of time dilatation and the invariance of length intervals, it generalises to
       $u = v+u'/\gamma_v$ where $u$ and $u'$ are the observed velocities of an object in S and S' respectively,
        {\it in the same space-time experiment}. Transposing this equation gives
       $u' = \gamma_v(u-v)$, which shows that the observed velocity of an object in S' is $\gamma_v$ times the relative
       velocity of the object and the frame S' in the frame S. Since the detector \DU in Figs. 5 and has
       $u = 0$ it follows that its speed in the frame S' is not $v$, as shown in Fig.~6, but instead $\gamma_v v$.
       What are actually shown in Fig.~6  are, not, as claimed in the paper, the same events shown in Fig. 5, but viewed
       in S', but rather the kinematical configurations of the {\it physically-independent reciprocal experiment}. 
       Indeed the distinct and different TD effects according to Eqn(2.11) and (2.13) are evident in Fig.~5c and 6c
       respectively.
       \par The `relativity of simultaneity' effects discussed in connection with Figs. 5, 6 and 7 are therefore spurious,
        since the times which are compared are not observations of the same events in different frames in the same
        space-time experiment, but rather observations of events in the three {\it physically-independent} experiments
       that are shown in the figures. In the primary photon-clock experiment shown in Fig. 5, clocks at rest in S' run slower
       than clocks at rest in S, whereas in the reciprocal experiment shown in Fig. 6 clocks at rest in S run slower than
       clocks at rest in S'.
        Because of the symmetry of the configurations in Fig. 7, clocks at rest in S and S' run slower, by the same amount, 
       as clocks at the rest in S''.
       \par The correct analysis of Einstein's TETE on the basis of the non-simultaneous events in S and S' (which
       appear nowhere in Figs.~5, 6 and 7) according to the RRVTR is presented in Ref.~\cite{JHFRECP}. These events
        transform into non-simultaneous events in the frame S', but with prior knowledge of the parameters
        $c$ and $v$ and of the RRVRT it may be deduced from the observed times of the non-simultaneous light signal events
        that the lightning-stroke events, that are the sources of the light signals, are simultaneous in the train frame S'
        contrary to Einstein's conclusion.
        \par With the exception of the assertion of the correctness of Nelson's Fig.~4 as corresponding configurations
         in the frames S and S' in the experiment ($v$ in (b) should be replaced by $\gamma_v v$) the criticisms
        of Ref.~\cite{Nelson} remain valid ---the assertion, in the latter, in agreement with Einstein, of non-simultaneity
        of the lightning-stroke events in the train frame is indeed erroneous.
       \par The same mistake concerning the physical interpretation of the postulate E2 as in Figs.~5-7 occurs in
       the analysis of the experiment presented in the Appendix. In Fig.~8b the world lines of A, A' and a light signal
       intersect in the same point as, simultaneously, do the world lines of  B, B' and another  light signal.
       As correctly stated in Section 2, the same coincidence events must be seen to occur, and to occur simultaneously,
      in all other inertial frames, in particular in that, S',  of the train. Inspection of Figs.~10 and 11 shows
      that this is not the case for the calculation presented in the Appendix where it is assumed, according to
       the naive interpretation of the postulate E2, that the light signals have the same speed $c$ in both
       S and S' in the same space-time experiment. Use of the RRVTR to correctly transform the photon 
       velocities into the frame S' gives, correctly, the two simultaneous triple world intersections,
       but implies that the observed angles of the light signals in S' are not correctly given by
      the kinematical transformation formula (A.3) used in the Appendix to calculate the
     directions of the light signals. This important point will be further discussed in a
      forthcoming article.
       \par The derivation  of the relation $L = L'$ of Eqn(4.2) is wrong because it assumes that events on the world
            lines of the spatially separated points A'and B' both transform according to the `generic' LT
             (2.1)-(2.4). Using the correct transformation equations, (6.2) and (6.3) for a synchronised clock at
             A' as well as (6.6) and (6.7) for one at B', and setting $L$ to $L'$ on the left sides of (6.2) and (6.6),
             gives, instead of (4.1),

      \begin{eqnarray}      
    (\Delta x')^2 -c^2  (\Delta t')^2 & = & (\Delta x)^2 -c^2  (\Delta \tau)^2 \nonumber \\
         &   & +2 \gamma_v L'(\Delta x-L)-2\Delta x L-2 \gamma_v v\Delta \tau+L^2 +(L')^2
      \end{eqnarray}
     Setting $\Delta t' = \Delta \tau = 0$ and $\Delta x = L$ in this equation leads not to the
     relation $L = L'$ but to the identity
    \begin{equation}
    \Delta x' \equiv L' = L'
    \end{equation}
     In fact the equality of $L$ and $L'$ is established in a straightforward manner from (6.2)
       on replacing $L$ by $L'$ on the left side:
 \begin{equation}
 x'_{A'}+\frac{L'}{2}  =  \gamma[x_{A'}+\frac{L}{2}-v\tau_A] = 0
  \end{equation}
     Since $L$ and $L'$ in (20) are independent of $v$, this equation is valid for all values
    of $v$, in particular for $v = 0$, $\gamma = 1$ and $x \rightarrow x'$:
   \begin{equation}
     x'_{A'}+\frac{L'}{2}  =  x'_{A'}+\frac{L}{2}
   \end{equation}
     or 
   \begin{equation}
     L'  = L
   \end{equation}
{\bf Appendix}
\renewcommand{\theequation}{A.\arabic{equation}}
\setcounter{equation}{0}
 \par In the analysis above of Einstein's TETE, light signals produced by lightning strokes hitting the
  embankment were considered, as viewed by observers on the embankment or in the moving train. In this
  appendix the `lightning strokes' are replaced light pulses, generated by a source in the embankment
   frame, that strike the embankment at points with the same $x$-coordinates as the ends of the moving train
 at this instant. The events corresponding to the detection of these light pulses, 
  ---analogous to the lightning stroke events in the Einstein TETE--- are analysed directly rather than
  the light signals produced by, say, reflection of the pulses at the embankment, as in the Einstein TETE.
    \par A pulsed laser source (PLS) produces two bunches of photons directed towards the points A and B
   on the embankment (Fig.8a). The timing of the laser pulses is such that they arrive at the points A and B
 at the instant that the ends A' and B' of the train are aligned with A and B (Fig.8b). With parameters
  $\theta$ and $L$ as defined in Fig.8, this implies that the photon pulses arrive simultaneously at A and B
 at the time $\tau = L/(2c\cos \theta)$, if they are emitted from the PLS at $\tau = 0$. The configurations
   shown in Fig.8 correspond $\theta = 60^{\circ}$ and $\beta_v = 0.5$. Applying Einstein's interpretation
   of his TETE to this experiment it would be concluded that the photon pulses are observed
   simultaneoustly in both the embankment or the train frames. It will now be demonstrated that this
   conclusion, which is correct if the photon pulses are replaced by massive objects obeying Newtonian
   kinematics, is no longer true in special relativity, whether or not massive objects or
   massless photons are considered. To demonstrate this, the following relativistic calculations will be
   performed assuming massive particles rather than photons. Newtonian kinematics and the case of photons
   are given by taking the $c \rightarrow \infty$ and $m \rightarrow 0$ limits, respectively, of
   the relativistic equations for particles with an arbitary Newtonian mass $m$.
 
\begin{figure}[htbp]
\begin{center}\hspace*{-0.5cm}\mbox{
\epsfysize15.0cm\epsffile{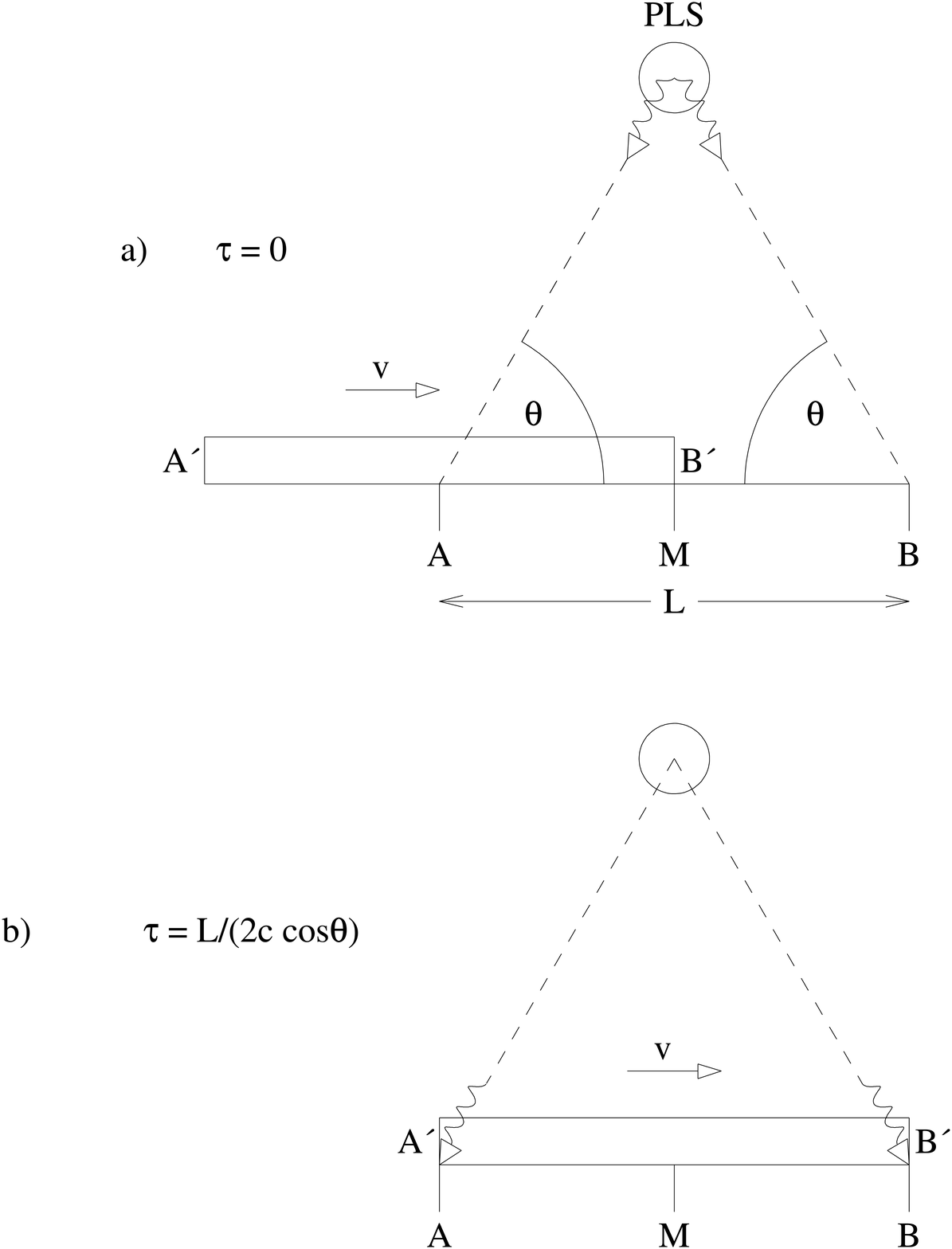}}
\caption{ {\em a) The pulsed laser source, PLS, produces, simultaneously, two bunches of
  photons directed towards the points A and B on the embankment at $\tau = 0$. b) at time
   $\tau = L/(2c \cos \theta)$ the photon bunches arrive at A and B at the instant that the
  points A' and B' on the train are $xt$-contiguous with A and B respectively. In this figure 
  $\beta_v = 0.5$.}}
\label{fig-fig8}
\end{center}
\end{figure}
\begin{figure}[htbp]
\begin{center}\hspace*{-0.5cm}\mbox{
\epsfysize15.0cm\epsffile{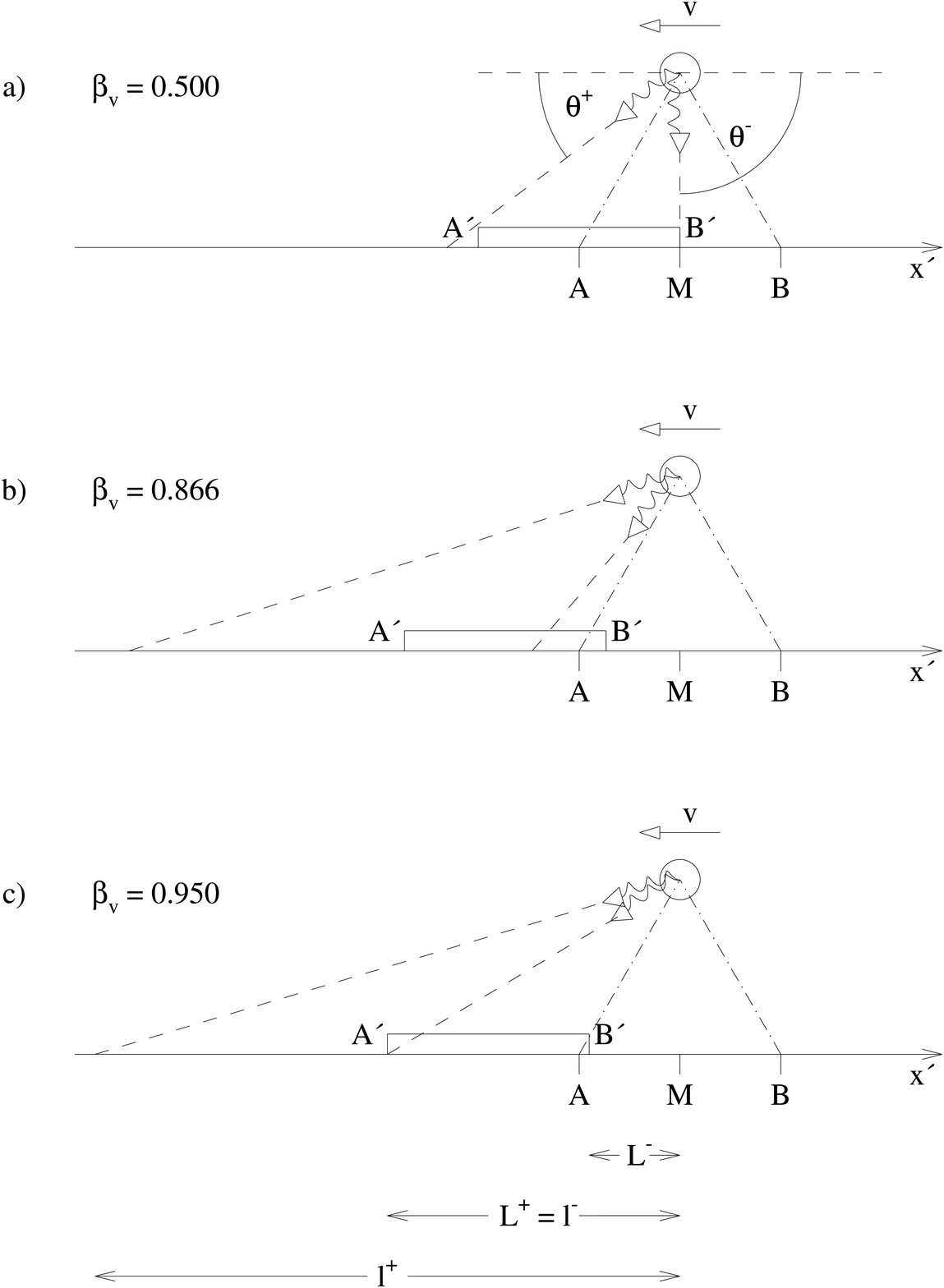}}
\caption{ {\em Directions and positions of the photon bunches from the PLS
   in the train frame at $\tau = \tau' = 0$ for various values of $\beta_v$.}}
\label{fig-fig9}
\end{center}
\end{figure}
\begin{figure}[htbp]
\begin{center}\hspace*{-0.5cm}\mbox{
\epsfysize15.0cm\epsffile{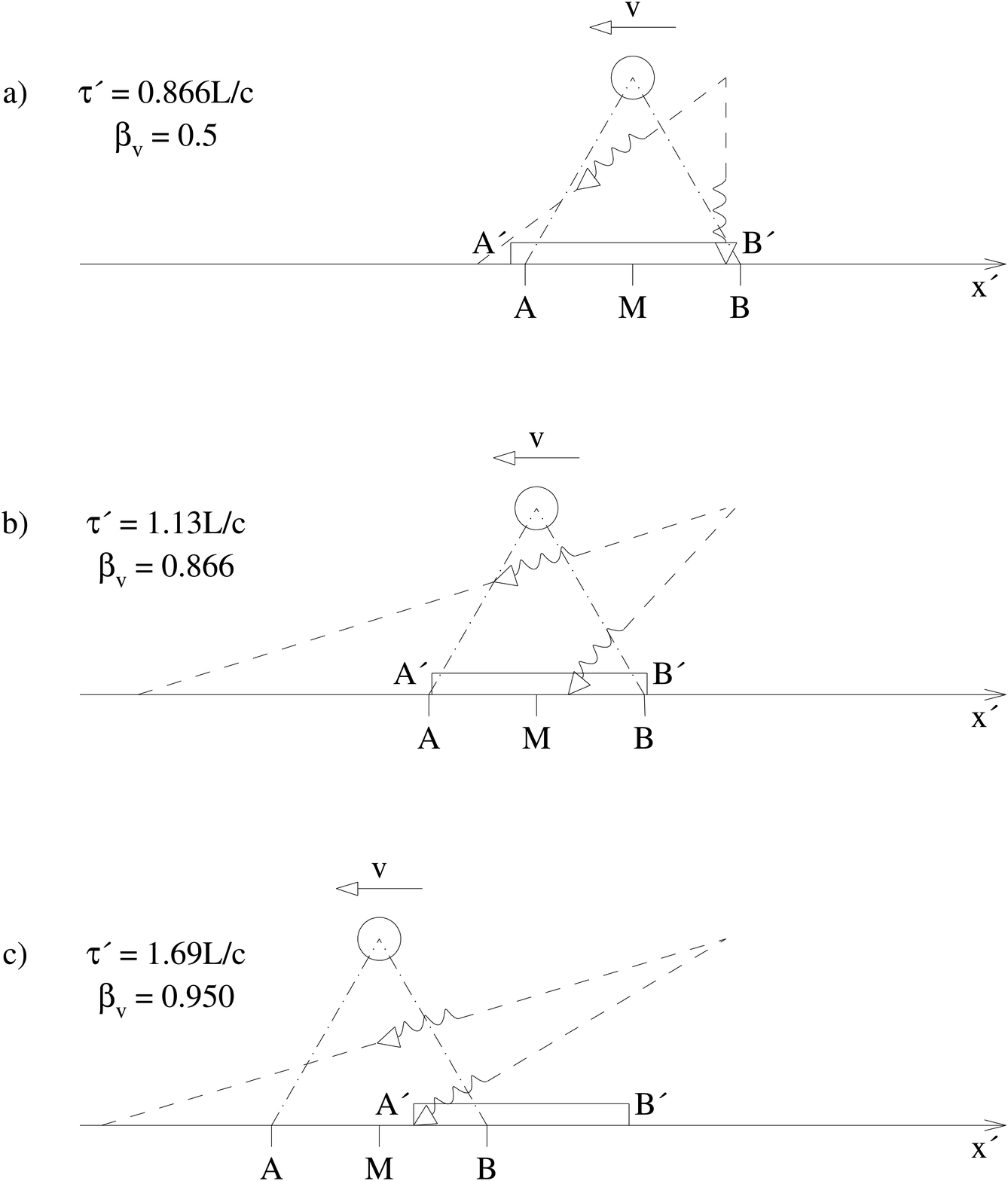}}
\caption{ {\em Positions of the photon bunches in the train frame at later times.
  a) $\beta_v = 0.5$, photon bunch $xt$-contiguous with B'; b) $\beta_v = 0.866$, photon bunch crosses $x'$-axis;
  c) $\beta_v = 0.950$, photon bunch $xt$-contiguous with A'.}}
\label{fig-fig10}
\end{center}
\end{figure}
\begin{figure}[htbp]
\begin{center}\hspace*{-0.5cm}\mbox{
\epsfysize10.0cm\epsffile{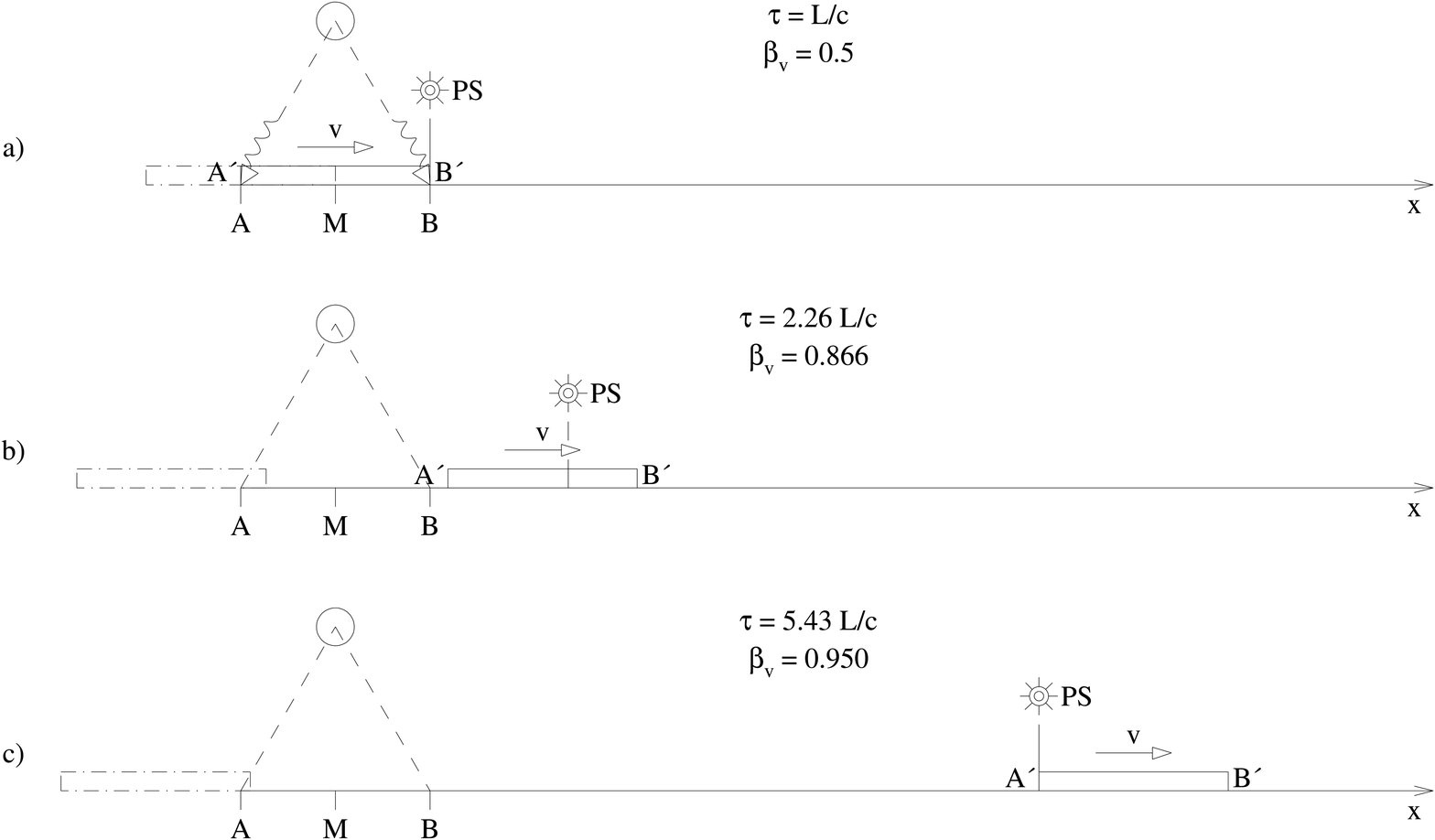}}
\caption{ {\em The train-frame  $xt$-contiguous events shown in Fig.10 as observed from the embankment frame. 
 Coincidences registered by detectors in the train produce the prompt signals, PS, shown. The
    positions of the train at $\tau= 0$ are shown dot-dashed.}}
\label{fig-fig11}
\end{center}
\end{figure}
\begin{figure}[htbp]
\begin{center}\hspace*{-0.5cm}\mbox{
\epsfysize15.0cm\epsffile{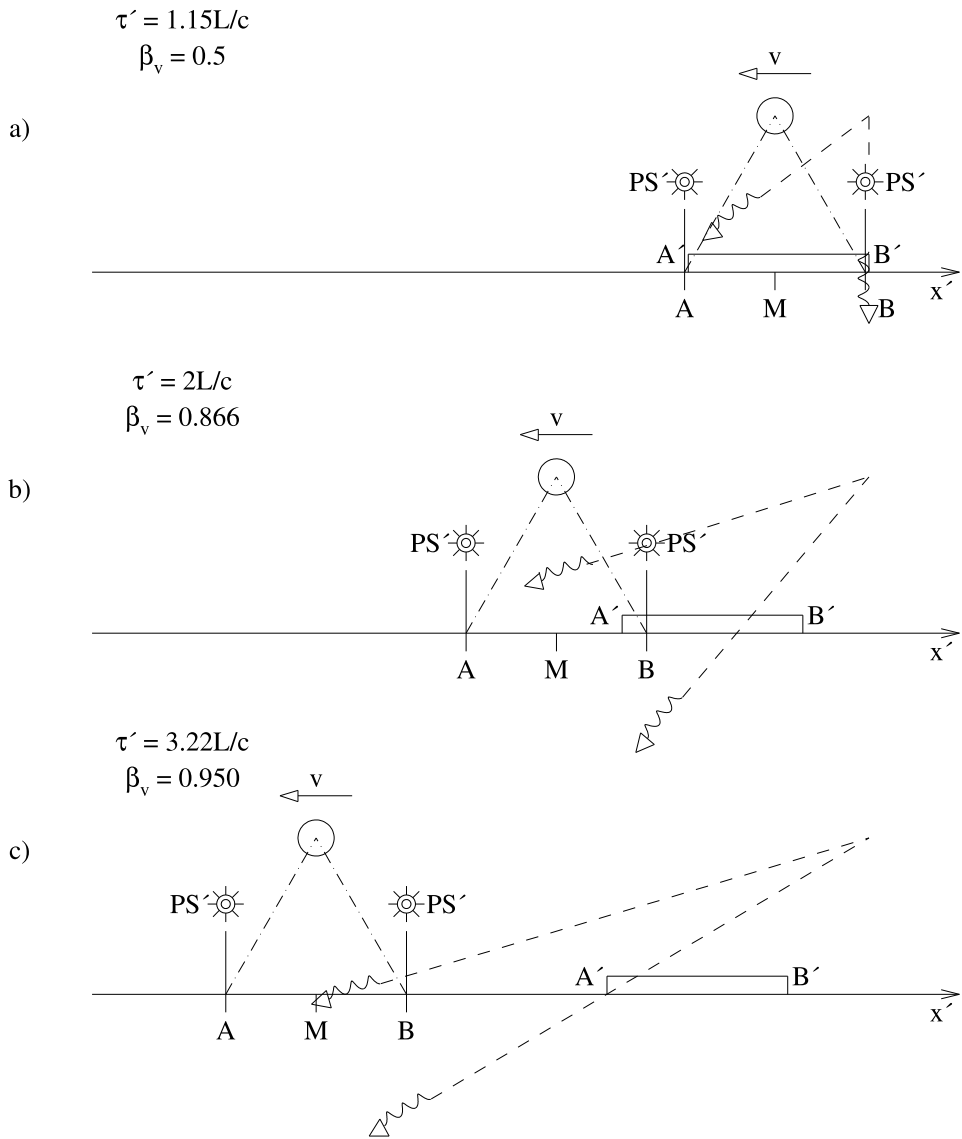}}
\caption{ {\em The simultaneous embankment frame photon-A and photon-B $xt$-contiguous events
    of Fig.8b, as viewed from the train frame, showing also the correponding positions
    of the photon bunches in this frame.  The  photon-A and photon-B coincidence events
  in the embankment frame produce
   prompt signals, PS', shown. These events are also simultaneous in the train frame. In this
   case there is no `relativity of simultaneity'.}}
\label{fig-fig12}
\end{center}
\end{figure}

   \par Transforming the photon momenta into the rest frame of the train at $\tau = \tau' = 0$ gives, for the photon case, the
        kinematical and spatial configurations shown in Fig.9 for $v = c/2$, $\sqrt{3}c/2$ and $0.950c$.
    The rationale for the choice of the last value of $v$ is explained  below. Shown in Fig.9, in each case,
     are the positions of the train and of the points A, M and B on the embankment defined in the same way
     as in Einstein's TETE, as well as the positions and directions of flight of the photon bunches
      \footnote{ In Figs. 8a and 9, the the photon bunches are located at the tails of the wavy arrows,
       in Figs.8b and 10-12 at the tips of the arrows.}. 
           
      The latter are derived from the LT of relativistic momentum according to the equations:
       \begin{eqnarray}
       p^{\pm} \cos \theta^{\pm} & = & \gamma_v ( p \cos \theta \pm \frac{\beta_v E}{c})
        \\
      p^{\pm} \sin \theta^{\pm} & = & p \sin \theta 
        \end{eqnarray}
      Eqn(A.1) is the transformation equation for the longitudinal component of momentum, while Eqn(A.2)
      expresses conservation of transverse momentum.
       Eqns(A.1) and (A.2), together with the relations $E = \gamma_u m c^2$ and $p = \gamma_u \beta_u m c$,
        where the particle of mass $m$ is assumed to move with speed $u$, then give:
       \begin{equation}
       \tan \theta^{\pm} = \frac{u \sin \theta}{\gamma_v (u \cos \theta \pm v)}
       \end{equation}
          The angles $\theta $ and $\theta^{\pm}$ are defined on Fig.8 and Fig.9a respectively. 
    If $L^+$ and  $L^-$ are the (positive) distances of A' and B', respectively, from M and $l^{\pm}$
       are the (positive) distances from M of the intersections of the particle trajectories
        with the $x'$-axis (see Fig.9c), then the geometry of Fig.9 gives:
         \begin{equation} 
         L^{\pm} = \frac{L}{2}\left(\frac{v}{u \cos \theta} \pm 1  \right)      
        \end{equation} 
    and
          \begin{equation} 
         l^{\pm} = \frac{\gamma_v L}{2}\left(\frac{v}{u \cos \theta} \pm 1  \right)      
         \end{equation}
      It follows from (A.4) and (A.5) that:
           \begin{equation}
          l^{\pm} =  \gamma_v  L^{\pm}
            \end{equation}
    In the limit $c \rightarrow \infty$, $\gamma_v \rightarrow 1$ of Galilean relativity, $ l^{\pm} = L^{\pm}$,
    so the particle trajectories pass through the points A' and B' at the end of the train for any values,
     $ \ll c$, of $v$ and $u$. In this limit (A.2) becomes:
     \begin{equation}
    u^{\pm} \sin \theta^{\pm} = u \sin \theta  
    \end{equation}
      showing that both particles have the same transverse velocity. They then arrive simultaneously
     at A' and B' in both the train and the embankment frames.
      \par As is clear from inspection of Fig.9, this is no longer the case for photons, given by setting
       $u = c$ in (A.4) and (A.5). All the configurations in Fig.9 have $\theta = 60^{\circ}$. Fig.9a
     corresponds to $\cos \theta = \beta_v = 1/2$, so that $L^{-} = l^{-} = 0$ and a photon 
     trajectory passes through B'. In general, however, (see Fig.9b) the photon trajectories do not 
     pass through A' and B', so contrary to the naive (classical) expectation drawn from inspection of Fig.8b,
     detectors situated at A' and B' would register no photon detection events.
      \par In Fig.9c, $v$ is chosen so that the condition $L^{+} = l^{-}$ is respected. In this case the 
     photon trajectory at the larger angle relative to the $x'$-axis passes through A'. Setting
      $L^{+} = l^{-}$, $u = c$ and $\cos \theta = 1/2$  in (A.4) and (A.5) gives a cubic equation for
       $\beta_v$:
         \begin{equation}
       4 \beta_v^2(1+\beta_v)+\beta_v-8 = 0 
        \end{equation}
    The solution of this equation is $\beta_v = 0.950$ or $\gamma_v = 3.22$. 
      \par In Fig.10 are shown the configurations in the train frame (positions of
     the embankment and the photon bunches) for the three cases shown in Fig.9, when the 
     photon bunches with largest angles to the $x'$-axis cross the latter. It can be seen that in
     no case, contrary to what is observed in the embankment frame, are the points A,A' or B,B'
     contiguous when the photon trajectories pass through A' or B'.    
     If it is now imagined that photon detectors equipped with signal sources similar
     to Y', \YU, \YL and Y'' in Fig.4, register photon coincidences in the train for the configurations
      shown in Fig.10, the corresponding signals (calculated using the TD formula (2.11)), as observed
    in the embankment frame, are shown in Fig.11. Only for $\beta_v = \cos \theta = 1/2$ are the
    photon-B' coincidence in the train frame and and the photon-B coincidence in the embankment frame
   observed simultaneously in the latter. This is easy to understand as the consequence, in this
     case, of exact compensation  between the shorter path length in S' tending to
    make the photon-B' coincidence earlier than the photon-B one, and the TD
     effect that tends to make it later.
     Finally, in Fig.12 are shown the simultanous photon-A and photon-B coincidence events of Fig.8b,
    as viewed from the train frame, for different values of $\beta_v$ In all cases these events
    are also simultaneous in the train frame. However, following Einstein's methodology in the orginal
    TETE and placing an observer at the middle of the train, this observer sees always the 
     photon-B coincidence events before the photon-A ones due to different light propagation time
     delays. The observer will then judge, by Einstein's criterion of simultaneity, that the events
    are not simultaneous, whereas, in fact, they are.

   \par It is interesting to note the huge differences between the AA' and BB' separations
    on the embankment and in the train in Figs.10, 11 and 12, for corresponding event configurations in the train
    and on the embankment, when $\beta_v$ is of order unity.
    This is the most deeply counter-intuitive effect among all the predictions for space-time experiments given by
    special relativity theory.

\pagebreak

\end{document}